\documentclass[preprint,aps,nofootinbib]{revtex4-1}
\pdfoutput=1
\usepackage{bm,amsmath,amssymb,amsfonts,slashed,array,graphicx,soul}
\usepackage[vcentermath,noautoscale]{youngtab}
\usepackage{mathrsfs}
\usepackage{xcolor}
\usepackage{hyperref}
\usepackage{cancel}
\usepackage[normalem]{ulem}
\usepackage{slashed}            
\usepackage{tikz}
\usetikzlibrary{arrows,shapes}
\usetikzlibrary{trees}
\usetikzlibrary{matrix,arrows} 				
\usetikzlibrary{positioning}				
\usetikzlibrary{calc,through}				
\usetikzlibrary{decorations.pathreplacing}  
\usetikzlibrary{decorations.pathmorphing}	
\usetikzlibrary{decorations.markings}
\usetikzlibrary{snakes}
\usepackage[normalem]{ulem}
\tikzset{
    vector/.style={decorate, decoration={snake}, draw},
	provector/.style={decorate, decoration={snake,amplitude=2.5pt}, draw},
	antivector/.style={decorate, decoration={snake,amplitude=-2.5pt}, draw},
    fermion/.style={draw, postaction={decorate},
        decoration={markings,mark=at position .55 with {\arrow[draw]{>}}}},
    fermionbar/.style={draw, postaction={decorate},
        decoration={markings,mark=at position .55 with {\arrow[draw=black]{<}}}},
    fermionnoarrow/.style={draw},
    gluon/.style={decorate, draw,decoration={coil,amplitude=4pt, segment length=6pt}, line width=1},
    scalar/.style={dashed,draw, postaction={decorate},
        decoration={markings,mark=at position .55 with {\arrow[draw]{>}}}},
    scalarbar/.style={dashed,draw, postaction={decorate},
        decoration={markings,mark=at position .55 with {\arrow[draw]{<}}}},
    scalarnoarrow/.style={dash pattern = on 6 pt off 3 pt,draw},
    electron/.style={draw, postaction={decorate},
        decoration={markings,mark=at position .55 with {\arrow[draw]{>}}}},
	bigvector/.style={decorate, decoration={snake,amplitude=4pt}, draw},
	vectorscalar/.style={loosely dotted,draw, postaction={decorate}},
}

\def\lsim{\mathrel{\rlap{\lower4pt\hbox{\hskip1pt$\sim$}}
    \raise1pt\hbox{$<$}}}
\def\gsim{\mathrel{\rlap{\lower4pt\hbox{\hskip1pt$\sim$}}
    \raise1pt\hbox{$>$}}}
\renewcommand{\thefootnote}{\fnsymbol{footnote}}
\begin{document}
\title{Charged Composite Scalar Dark Matter}
\author{Reuven Balkin, Maximilian Ruhdorfer, Ennio Salvioni, and Andreas Weiler$^{\;}$}\email[Email: ]{reuven.balkin@tum.de}\email{max.ruhdorfer@tum.de}\email{ennio.salvioni@tum.de}\email{andreas.weiler@tum.de}
\affiliation{Physik-Department, Technische Universit\"at M\"unchen, 85748 Garching, Germany}
\begin{abstract} We consider a composite model where both the Higgs and a complex scalar $\chi$, which is the dark matter (DM) candidate, arise as light pseudo Nambu-Goldstone bosons (pNGBs) from a strongly coupled sector with TeV scale confinement. The global symmetry structure is $SO(7)/SO(6)$, and the DM is charged under an exact $U(1)_{\rm DM} \subset SO(6)$ that ensures its stability. Depending on whether the $\chi$ shift symmetry is respected or broken by the coupling of the top quark to the strong sector, the DM can be much lighter than the Higgs or have a weak-scale mass. Here we focus primarily on the latter possibility. We introduce the lowest-lying composite resonances and impose calculability of the scalar potential via generalized Weinberg sum rules. Compared to previous analyses of pNGB DM, the computation of the relic density is improved by fully accounting for the effects of the fermionic top partners. This plays a crucial role in relaxing the tension with the current DM direct detection constraints. The spectrum of resonances contains exotic top partners charged under the $U(1)_{\rm DM}$, whose LHC phenomenology is analyzed. We identify a region of parameters with $f = 1.4$ TeV and $200\;\mathrm{GeV} \lesssim m_\chi \lesssim 400$ GeV that satisfies all existing bounds. This DM candidate will be tested by XENON1T in the near future. 

\end{abstract}
\preprint{TUM-HEP-1090-17}
\maketitle
\tableofcontents
\renewcommand{\thefootnote}{\arabic{footnote}}
\section{Introduction}
Models where the Higgs emerges as a composite field from a new, strongly interacting theory provide an attractive solution to the naturalness problem (see Refs.~\cite{Bellazzini:2014yua,Panico:2015jxa} for recent reviews). A key ingredient is that the strong sector is endowed with a global symmetry $\mathcal{G}$, spontaneously broken to the subgroup $\mathcal{H}$ at a scale $f \sim \mathrm{TeV}$. The four components of the Higgs doublet $H$ are assumed to be among the Nambu-Goldstone bosons that parameterize the $\mathcal{G}/\mathcal{H}$ coset, thus explaining the hierarchy between the mass of the Higgs and those of the other composite particles. To generate realistic couplings of $H$ to the Standard Model (SM) gauge bosons and fermions, some amount of explicit $\mathcal{G}$ breaking is required. This in turn generates a radiative potential for the Higgs that triggers electroweak symmetry breaking (EWSB).

The minimal and best-known model in this class \cite{Agashe:2004rs} is based on the $SO(5)/SO(4)$ coset, which is fully parameterized by the four components of $H$. Less minimal realizations, however, are equally plausible from a bottom-up perspective, and open up an interesting potential connection with the dark matter (DM) puzzle: If $\mathcal{G}/\mathcal{H}$ contains additional pseudo Nambu-Goldstone bosons (pNGBs) beyond $H$, one or more of these may be stable on cosmological time scales, and serve as cold DM candidate(s). This simple idea is very appealing, because the pNGB nature of the scalar DM would naturally explain why it is light and weakly coupled to the SM at low energies. Furthermore, it is highly predictive, since the DM interactions and mass are determined by the global symmetry structure and by the few parameters (spurions) that break it explicitly. The latter are at least in part fixed by the requirement of reproducing the measured SM parameters, such as the Higgs vacuum expectation value (VEV) and mass, and the top Yukawa coupling. This setup was first explored in Ref.~\cite{Frigerio:2012uc}, where an additional real pNGB $\eta$ was considered as DM candidate, and its phenomenology was analyzed within the low-energy effective field theory (EFT) of the $SO(6)/SO(5)$ model \cite{Gripaios:2009pe}, which yields $(H, \eta)$ as Goldstone bosons. It was pointed out that higher-dimensional derivative operators play a central role in the phenomenology, setting composite pNGB DM apart from the well-known renormalizable Higgs portal model \cite{Silveira:1985rk,McDonald:1993ex,Burgess:2000yq}. Subsequently, Ref.~\cite{Marzocca:2014msa} performed an extensive phenomenological analysis of the $SO(6)/SO(5)$ model, introducing explicitly the lightest composite resonances and imposing generalized Weinberg sum rules (WSRs) \cite{Weinberg:1967kj} to render the scalar potential for $H$ and $\eta$ calculable \cite{Marzocca:2012zn,Pomarol:2012qf}. 

In the $SO(6)/SO(5)$ model, the DM can be stabilized by an exact $Z_2$ symmetry \mbox{$P_\eta:\; \eta \to - \eta\,$}. Even though the leading two-derivative Lagrangian preserves this symmetry, this is not automatically the case for higher order terms, because \mbox{$P_\eta \notin SO(6)$}. For example, at the four-derivative order a Wess-Zumino-Witten (WZW) term appears, $\sim c_W \eta (g^2 W_{\mu\nu}^a\tilde{W}^{a\,\mu\nu} - g^{\prime\,2} B_{\mu\nu}\tilde{B}^{\mu\nu})/(16\pi^2 f)$, where $c_W$ is an anomaly coefficient that depends on the ultraviolet (UV) completion \cite{Gripaios:2009pe}. Thus, to ensure DM stability one is forced to assume that the UV theory respects the extended symmetry $O(6)/O(5)$, which guarantees that $P_\eta \in O(5)$ is respected at all orders, and sets in particular $c_W = 0$.

In this paper we consider a different DM stabilization mechanism: The DM is a complex scalar, charged under an exact $U(1)_{\rm DM}$ which is a global or gauged subgroup of the unbroken symmetry $\mathcal{H}$. This mechanism is UV-robust, in the sense that any $\mathcal{G}$-invariant high energy completion describing the strong sector will automatically preserve the $U(1)_{\rm DM}$ and therefore the DM stability. For a concrete realization we adopt the $SO(7)/SO(6)$ coset, which contains in addition to $H$ two real Goldstones $\eta, \kappa$. The unbroken $SO(6)$ possesses an $SO(4) \times SO(2)$ subgroup, with $H$ transforming as a $\mathbf{4}$ under $SO(4)\cong SU(2)_L \times SU(2)_R$, whereas the $SO(2)$ rotates $\eta$ and $\kappa$. The $SO(2)$ is manifestly preserved by the SM gauge interactions and, as we will see, can also be respected by the couplings of the SM fermions to the strong sector. In this case $\eta$ and $\kappa$ form a complex scalar field, charged under the exact $SO(2)\cong U(1)_{\rm DM}\,$. This complex pNGB scalar, dubbed $\chi$, is the DM candidate. 

In Sec.~\ref{sec:DMstability}, we will discuss the predicted DM mass ranges. Depending on whether the couplings of the top quark do or do not break the shift symmetry protecting $\chi$, the DM mass can be naturally of $O(\mathrm{few})\times 100\, \mathrm{GeV}$ or much smaller than the Higgs mass. These two options lead to very distinct phenomenology. In this paper we focus on the scenario where $m_\chi$ is at the weak scale, whereas the case of light DM will be the subject of a separate publication \cite{LightDMfuture}. In the spirit of Ref.~\cite{Marzocca:2014msa}, we introduce an effective Lagrangian for the strong sector resonances and obtain calculability of the potential for $H$ and $\chi$ by imposing a set of WSRs. Our working model contains two layers of fermionic resonances (the ``top partners'') and one layer of vector resonances. We perform a detailed phenomenological analysis, considering the constraints from the observed value of the DM relic abundance, as well as from direct detection experiments and from the LHC. Indirect detection is also briefly discussed. As first pointed out in Ref.~\cite{Frigerio:2012uc}, the DM phenomenology is characterized by the interplay of higher-dimensional derivative operators, which are important in annihilation but suppressed by the tiny momentum transfer in the scattering off nuclei, with the marginal (radiative) portal coupling $\sim \lambda \chi^\ast \chi H^\dagger H$, which is energy-independent.

We identify two effects that were not considered in previous analyses, but can change significantly the prediction for the pNGB DM relic abundance. Firstly, in Sec.~\ref{sec:annEFT} we show that the mixing between the top and its partners reduces the $t\bar{t}\chi^\ast \chi$ coupling. Somewhat counter-intuitively (see Sec.~\ref{subsec:DMra} for the details), this turns out to suppress the size of the portal coupling $\lambda$ necessary to reproduce the relic abundance, relaxing the tension with direct detection experiments. This is, in fact, essential to evade the current constraints from XENON1T~\cite{Aprile:2017iyp} and LUX~\cite{Akerib:2016vxi}. Secondly, we find that the derivative operators receive large, only partially calculable radiative corrections proportional to the explicit $\mathcal{G}$-breaking in the top sector, discussed in Sec.~\ref{sec:finitemom}. These radiative corrections imply an irreducible theoretical uncertainty on the annihilation cross section, estimated at $50\%$, which broadens the viable region of parameter space. The outcome of our analysis is that there exists a region of parameters, with DM mass in the $200\,$-$\,400$ GeV range, that is compatible with all current constraints but will be fully tested by direct detection experiments like XENON1T in the near future. We discuss the LHC exclusions and future prospects on the fermionic top partners, some of which are charged under the $U(1)_{\rm DM}$ and therefore decay into a top quark and a DM particle, yielding ``stop-like'' signatures. The fine-tuning associated to the theory is also estimated. Notice that throughout the discussion we will assume that $U(1)_{\rm DM}$ is a global symmetry. A few comments about the consequences of gauging it are given in the Outlook, Sec.~\ref{sec:outlook}.

The $SO(7)/SO(6)$ coset has been recently discussed in Ref.~\cite{Chala:2016ykx}, taking as DM candidate the real scalar $\eta$, stabilized by a $Z_2$ symmetry which is not an element of $SO(7)$. While $SO(7)$ does not admit complex representations, implying that a WZW term is absent,\footnote{See Ref.~\cite{Chala:2012af} for earlier remarks.} in general other higher-derivative operators may break the parity, so the assumption still needs to be made that an $SO(7)$-invariant UV completion exactly preserves it. Furthermore, Ref.~\cite{Chala:2016ykx} aimed at incorporating electroweak baryogenesis and concentrated on a region of parameters that does not overlap with ours. See also Refs.~\cite{Barnard:2014tla,Ma:2017vzm,Ballesteros:2017xeg} for other composite Higgs models with pNGB DM candidates.

The remainder of the paper is organized as follows. In Sec.~\ref{sec:model} we present the essential ingredients of the model, discuss the DM stabilization and introduce the composite resonances. In Sec.~\ref{sec:scalarpotential} the scalar potential is studied and the WSRs that render it calculable are implemented. In addition, a first characterization of the realistic parameter space is performed. Section~\ref{sec:pheno} is dedicated to the DM phenomenology: We present our calculation of the DM relic abundance, including the novel effects previously mentioned, then we discuss the direct detection constraints and comment on indirect detection. Our main results are presented in Sec.~\ref{sec:results}. In Sec.~\ref{sec:LHC} we analyze the collider phenomenology of the model. Finally, our concluding remarks are offered in Sec.~\ref{sec:outlook}. The technical aspects of the analysis are summarized in three Appendices: App.~\ref{app:A} contains the details of the model, App.~\ref{app:B} presents the complete scalar potential as well as our procedure for the parameter scan, and App.~\ref{app:C} collects some formulas for the DM phenomenology that were omitted from the main text.

\section{SO(7)/SO(6) Model} \label{sec:model}
We assume that the strong sector possesses an $SO(7)$ global symmetry, spontaneously broken to $SO(6)$ at the scale $f$. The six Goldstone bosons (GBs) $\pi^{a},\ a=1,\ldots ,6$ transform in the fundamental representation of the unbroken $SO(6)$, which under $SO(4)$ decomposes into $H\sim \mathbf{4}$, identified with the Higgs doublet, and two real singlets $\eta,\kappa$. Following the Callan-Coleman-Wess-Zumino (CCWZ) construction \cite{Coleman:1969sm,Callan:1969sn}, whose details are given in App.~\ref{app:A}, the GBs are parameterized by the matrix $U= \exp \left ( i \sqrt{2} \pi^{a} X^{a}/f\right)$, where the $X^{a}$ are the broken generators. At the leading order in derivatives, the Goldstone Lagrangian is given by
\begin{equation} \label{eq:2deriv}
\mathcal{L}_{\pi} = \frac{f^2}{4} d_\mu^{a} d^{{a}\,\mu},
\end{equation}
where the CCWZ $d_\mu$ symbol is constructed out of $U$ and its $SU(2)_L\times U(1)_Y$ covariant derivative. In the unitary gauge the vector of GBs can be written as
\begin{equation} \label{eq:pionsUG}
\vec{\pi}=\left(0,\ 0,\ 0,\ \tilde{h},\ \eta,\ \kappa\right)^T,
\end{equation}
with $\tilde{h}$ denoting the field whose physical excitation will be identified with the observed Higgs boson. After performing a convenient field redefinition (see Eq.~\eqref{eq:fieldRedef}), the Goldstone Lagrangian reads 
\begin{equation}
\mathcal{L}_{\pi} = \frac{1}{2}\Big[(\partial_\mu \tilde{h})^2 + (\partial_\mu \eta)^2 + (\partial_\mu \kappa)^2\Big] + \frac{1}{2}\frac{\big(\tilde{h}\partial_\mu \tilde{h} +	\eta\partial_\mu\eta + \kappa\partial_\mu\kappa\big)^2}{f^2- {\tilde{h}}^2-\eta^2-\kappa^2}\\
+ \frac{\tilde{h}^2}{4}\Big[\bar{g}^2 |\bar{W}_\mu^+|^2 +\frac{1}{2}\big(\bar{g}\bar{W}_\mu^3 - \bar{g}'\bar{B}_\mu\big)^2 \Big].
\label{eq:GBLagrangian}
\end{equation}
The bar on the gauge fields (and their associated couplings) indicates that these are elementary states. In analogy to photon-rho mixing in QCD, the gauge fields couple linearly to resonances of the strong sector. The resulting mass mixing is diagonalized, for example for the charged fields, by $\bar{g} \bar{W}_\mu^\pm \rightarrow g W_\mu^\pm + \ldots\,$, where $g$ and $W^\pm_\mu$ are the SM gauge coupling and field, respectively, and the dots stand for terms containing the vector resonances (see Eq.~\eqref{eq:vectorMix}). Hence we identify $\langle \tilde{h} \rangle = v \simeq 246$ GeV. Assuming furthermore $\left\langle \eta \right\rangle = \left\langle \kappa \right\rangle = 0$ and expanding around the vacuum, we find that the singlets have canonical kinetic terms, whereas for the Higgs the canonical normalization is achieved with
\begin{equation} \label{eq:canonicalh}
\tilde{h} = v + \sqrt{1-\xi}\ h\,,\qquad  \xi \equiv \frac{v^2}{f^2}\,,
\end{equation}
where $h$ is the physical excitation.

\subsection{Coupling to elementary fermions and dark matter stability} \label{sec:DMstability}
We assume partial compositeness to be realized also in the fermionic sector \cite{Kaplan:1991dc}, where the elementary fermions are coupled linearly to operators of the strong dynamics. For the top quark we have schematically
\begin{equation} \label{eq:PCfermions}
\mathcal{L}_{\rm mix} \sim \epsilon_q\, \bar{q}_L \mathcal{O}_q + \epsilon_t\, \bar{t}_R \mathcal{O}_t + \mathrm{h.c.}\,.
\end{equation}
Notice that to reproduce the hypercharge of the SM fermions it is necessary to enlarge the global symmetry pattern to $SO(7) \times U(1)_X \to SO(6) \times U(1)_X$, such that $Y = T_R^{3} + X$. The $SO(7)_{X}$ representations of the operators $\mathcal{O}_{q,\,t}$ are not uniquely fixed. Since the SM fermions do not fill complete multiplets of $SO(7)$,\footnote{An exception is the case where $t_R$ is embedded into an $SO(7)$ singlet, see below.} the interactions in Eq.~\eqref{eq:PCfermions} break explicitly at least some of the shift symmetries protecting the GBs. A basic requirement is that the Higgs shift symmetry be broken, in order for $\tilde{h}$ to acquire a Yukawa coupling to the top and a $1$-loop potential of the correct size. In addition, the $SO(7)$ quantum numbers of $\mathcal{O}_{q,\,t}$ determine the structure of the couplings and potential for the singlets. To examine the different options for the quantum numbers, it is convenient to consider the decomposition of the representations of $SO(7)$ under its subgroup $SO(4) \times SO(3)$, which we can write as $SU(2)_L \times SU(2)_R \times SU(2)'$, where $SO(4) \cong SU(2)_L \times SU(2)_R $ while $SO(3) \cong  SU(2)'$ is generated by the broken generators under which the two singlets shift, $X^{\eta} \equiv X^5$ and $X^{\kappa} \equiv X^6$, together with 
\begin{equation}
T^{\rm DM} \equiv T^{56} = \frac{1}{\sqrt{2}}\, \mathrm{diag}(\mathbf{0}_{4\times 4}, \sigma^2, 0) \;\in\; SO(6),
\end{equation}
where we used block notation. The label given to this generator anticipates its role in the dark matter stabilization, which will be discussed momentarily. For the first few irreducible $SO(7)$ representations we have the following $(SU(2)_L, SU(2)_R, SU(2)')$ decompositions (see for example Ref.~\cite{Yamatsu:2015npn}),
\begin{equation}
\begin{split}
\mathbf{1} &\, =\, (\mathbf{1}, \mathbf{1}, \mathbf{1}),\\
\mathbf{7} &\, =\, (\mathbf{2}, \mathbf{2}, \mathbf{1}) \oplus (\mathbf{1}, \mathbf{1}, \mathbf{3}),\\
\mathbf{8} &\, =\, (\mathbf{2}, \mathbf{1}, \mathbf{2}) \oplus (\mathbf{1}, \mathbf{2}, \mathbf{2}),\\
\mathbf{21} &\, =\, (\mathbf{2}, \mathbf{2}, \mathbf{3}) \oplus (\mathbf{3}, \mathbf{1}, \mathbf{1}) \oplus (\mathbf{1}, \mathbf{3}, \mathbf{1}) \oplus (\mathbf{1}, \mathbf{1}, \mathbf{3}),  \\
\mathbf{27} &\, =\, (\mathbf{3}, \mathbf{3}, \mathbf{1}) \oplus (\mathbf{2}, \mathbf{2}, \mathbf{3}) \oplus (\mathbf{1}, \mathbf{1}, \mathbf{5}) \oplus (\mathbf{1}, \mathbf{1}, \mathbf{1}).
\end{split}
\label{eq:replist}
\end{equation}
To guarantee custodial protection against zero-momentum corrections to the $Zb_L \bar{b}_L$ vertex, which would conflict with LEP measurements, $q_L$ must be embedded in the $(\mathbf{2}, \mathbf{2})_{2/3}$ representation of $(SU(2)_L,\, SU(2)_R)_{X}$ \cite{Agashe:2006at}. Hence a natural and minimal possibility is $\mathcal{O}_q \sim \mathbf{7}_{2/3}$, which we adopt henceforth. In this case the coupling of $q_L$ preserves the entire $SU(2)'$. The requirement of $U(1)_X$-invariance of the top mass term fixes then the $X$ charge of $\mathcal{O}_t$ to be $2/3$, but several options are available for its transformation under $SO(7)$. 

\begin{figure}[t]
\centering
\includegraphics[scale=.75]{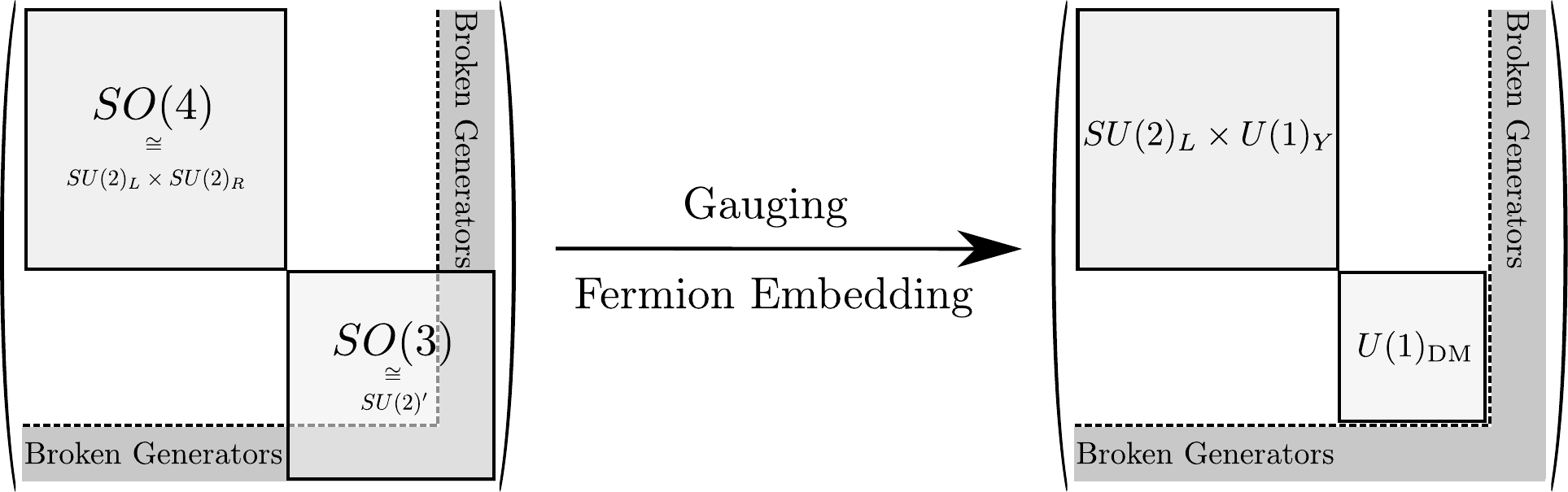}
\caption{Schematic overview of the $SO(7)$ algebra. In the left drawing the structure of the $SU(2)_L \times SU(2)_R \times SU(2)'$ subgroup is displayed, whereas the right drawing shows the symmetries that remain unbroken after the weak gauging of the SM electroweak group and the coupling of $q_L, t_R$ to operators $\mathcal{O}_q, \mathcal{O}_t \sim \mathbf{7}_{2/3}$ of $SO(7)_{X}$.}
\label{fig:GenBases}
\end{figure}

In this paper we focus on the choice
\begin{equation}
t_R \sim (\mathbf{1}, \mathbf{1}, \mathbf{3}) \subset \mathbf{7}\,.
\end{equation}
In this case the $SU(2)'$ is explicitly broken, but the embedding can be chosen as to preserve a residual $U(1)$, generated by one among $\{X^\eta, X^\kappa, T^{\rm DM}\}$. Therefore we can either leave the shift-symmetry of one of the singlets intact, thus keeping it massless, or preserve the $U(1)$ symmetry acting on $\eta$ and $\kappa$ that is generated by $T^{\rm DM}$. We choose the latter, hence $\eta$ and $\kappa$ are combined into a complex scalar field 
\begin{equation}
\chi \equiv \left(\kappa + i \eta\right)/\sqrt{2}\,
\end{equation}
that is an eigenstate of $U(1)_{\rm DM}$ with charge $+1$ (the normalization is fixed by taking $\sqrt{2} \,T^{\rm DM}$ as the generator), while $t_R$ is uncharged under this symmetry. The complex scalar $\chi$ is our DM candidate, and the unbroken $U(1)_{\rm DM}$ ensures its stability. This setup is represented schematically in Fig.~\ref{fig:GenBases}.
Under $(SU(2)_L, SU(2)_R)_{X}^{\rm DM}$ the $\mathbf{7}_{2/3}$ decomposes as
\begin{equation}
\mathbf{7}_{2/3} = (\mathbf{2},\mathbf{2})_{2/3}^0 \oplus (\mathbf{1},\mathbf{1})_{2/3}^0 \oplus (\mathbf{1},\mathbf{1})_{2/3}^{\pm 1}\,,
\end{equation}
where the $t_R$ is embedded in the $(\mathbf{1},\mathbf{1})_{2/3}^0$, while the $q_L$ is embedded in the $(\mathbf{2},\mathbf{2})_{2/3}^0$. The coupling of $t_R$ to the strong sector explicitly breaks the shift symmetry for $\chi$, which will acquire a potential, and in particular a mass, of the same parametric size as the Higgs. The explicit form of the embeddings is
\begin{equation} \label{eq:embed}
\xi_L = \frac{1}{\sqrt{2}}\begin{pmatrix}
i b_L, &  b_L, & i t_L, & - t_L, & 0, & 0, & 0 \end{pmatrix}^T\,,\;\qquad
\xi_R=\begin{pmatrix}
0, & 0, &  0, & 0, & 0, & 0, & t_R \end{pmatrix}^T\,.
\end{equation}

A different phenomenological scenario is realized if the embedding of $t_R$ preserves $SU(2)'$. Glancing at Eq.~\eqref{eq:replist}, this can be obtained in several ways: for example, $t_R\sim (\mathbf{1}, \mathbf{3}, \mathbf{1}) \subset \mathbf{21}$ (antisymmetric tensor) or $t_R\sim (\mathbf{1}, \mathbf{1}, \mathbf{1}) \subset \mathbf{27}$ (symmetric traceless tensor). Alternatively, we may assume that $t_R$ is a fully composite $SO(7)$ singlet. In all these cases the couplings of the top quark do not break the symmetries under which $\chi$ shifts, hence the leading contributions to its potential come from the couplings of the light fermions, from the weak gauging of $U(1)_{\rm DM}$, or from both. As a consequence, the DM is naturally much lighter than the Higgs. This intriguing possibility will be the subject of a separate publication \cite{LightDMfuture}. For $m_\chi < m_h/2$, an important constraint comes from the invisible decay width of the Higgs, mediated by the derivative interactions in Eq.~\eqref{eq:GBLagrangian}. The decay width is
\begin{equation}
\Gamma(h \to \chi^\ast \chi) = \frac{m_h^3 v^2}{16\pi f^4 (1 - \xi)} \sqrt{1 - \frac{4m_\chi^2}{m_h^2}}\,,
\end{equation}
where we neglected the contribution of the radiative portal coupling $\lambda$ (see Eq.~\eqref{eq:GBPotential} below), which is expected to be very small in the light $\chi$ scenario. The current $95\%$ CL lower bound of $\mathrm{BR}(h\to \chi^\ast \chi) < 0.24$ \cite{Khachatryan:2016whc} translates into $f \gtrsim 1.2$ TeV.  

One further comment about the DM stability is in order. In the above discussion we have assumed that each elementary fermion multiplet couples, in a $U(1)_{\rm DM}$-invariant way, to only one operator of the strong sector. However, in general additional, subleading couplings to other operators could be present. If any of these break the $U(1)_{\rm DM}$, the DM stability may be compromised. Therefore we need to make the assumption that the $U(1)_{\rm DM}$ is either a global symmetry respected by all elementary-composite mixing couplings, or an unbroken gauge symmetry.

\subsection{Resonances}\label{subs:resonances}
The strong sector resonances fill multiplets of the unbroken $SO(6)$ and can be consistently described in the CCWZ framework. We begin with the fermion sector, which plays a dominant role in our discussion. Since we have chosen to embed the third generation fermions in the fundamental of $SO(7)$, which decomposes as $\mathbf{7}_{2/3}=\mathbf{6}_{2/3}\oplus\mathbf{1}_{2/3}$ under $SO(6) \times U(1)_X$, we consider top partners in the fundamental $Q$ and singlet $S$ representations of $SO(6)$. The explicit expression of the fundamental is
\begin{equation}
Q=\frac{1}{\sqrt{2}}\begin{pmatrix}
i B - i X_{5/3}, & B + X_{5/3}, & iT + i X_{2/3}, & -T + X_{2/3}, & -i\mathcal{Y} + i \mathcal{Z}, &  \mathcal{Y} + \mathcal{Z} \end{pmatrix}^T\,.
\end{equation}
The doublet $(T,B)^T$ transforms as $\mathbf{2}_{1/6}^0$ under $\left(SU(2)_L\right)_{Y}^{\rm DM}$, and therefore has the same quantum numbers as $q_L$, whereas the exotic doublet $(X_{5/3},X_{2/3})^T \sim \mathbf{2}_{7/6}^0$ contains an exotic fermion with electric charge equal to $5/3$. The two states $\mathcal{Y},\mathcal{Z}\sim \mathbf{1}_{2/3}^{\pm 1}$ share the SM quantum numbers of the $t_R$, but are additionally charged under $U(1)_{\rm DM}$. The latter symmetry, being exact, strongly constrains their couplings. Finally, the quantum numbers of the $SO(6)$ singlet are $S\sim \mathbf{1}_{2/3}^0\,$. The leading order Lagrangian describing the fermion sector is
\begin{equation}
\begin{split}
\mathcal{L}_{f}\,&=\,\,i \bar{q}_L \slashed{D} q_L + i \bar{t}_R \slashed{D} t_R + \sum_{i=1}^{N_Q} \bar{Q}_i \left(i \slashed{D} + \slashed{e} - m_{Q_i}\right)Q_i + \sum_{j=1}^{N_S} \bar{S}_j \left(i \slashed{D} - m_{S_j} \right)S_j\\
& +\sum_{i=1}^{N_Q} \left(\epsilon_{tQ}^i \bar{\xi}^A_R U_{Aa} Q^{\,a}_{iL} + \epsilon_{qQ}^i \bar{\xi}_L^A U_{Aa} Q_{iR}^{\,a} \right)+ \sum_{j=1}^{N_S}  \left(\epsilon_{tS}^j \bar{\xi}_R^A U_{A7} S_{jL} + \epsilon_{qS}^j \bar{\xi}_L^A U_{A7} S_{jR} \right) + \mathrm{h.c.},
\end{split}
\label{eq:fermion}
\end{equation}
where $N_Q$ and $N_S$ denote the number of copies of each species of resonance that lie below the cutoff of the low-energy theory, and $A\,(a)$ is an index in the fundamental of $SO(7)\,(SO(6))$. The second line of Eq.~\eqref{eq:fermion} is the low-energy interpolation of Eq.~\eqref{eq:PCfermions}: the embeddings $\xi_{L,R}$ defined in Eq.~\eqref{eq:embed}, which transform linearly under $SO(7)$, have been `dressed' into reducible $SO(6)$ representations via insertions of the Goldstone matrix $U$. Also notice that the kinetic term of the $Q_i$ includes the $e_\mu$ symbol, which is necessary to respect the nonlinearly realized $SO(7)$. In general the following term should also be added to the Lagrangian,
\begin{equation}\label{eq:Fermint}
\mathcal{L}_{d} =  \sum_{i=1}^{N_Q} \sum_{j = 1}^{N_S}  \, c^L_{ji} \bar{S}_{jL} \,\slashed{d}^{\,a} Q_{iL}^{\,a} + \mathrm{h.c.} + (L \rightarrow R),
\end{equation}
where $c^{L,R}_{ji}$ are coefficients of $O(1)$. The operators in Eq.~\eqref{eq:Fermint} arise purely from the strong dynamics, and as a consequence they do not contribute to the scalar potential. At leading order in the $1/f$ expansion, they give rise to derivative interactions of one GB and two fermions, which scale as $\sim c^{L,R}\, p/f$, where $p$ is the relevant energy. In the processes relevant for DM phenomenology, namely annihilation and scattering with heavy nuclei, we have $p/f \lesssim m_\chi/f \ll 1$, hence these interactions are suppressed compared to the $\mathcal{G}$-breaking couplings that arise from Eq.~\eqref{eq:fermion}, which scale as $\sim \epsilon/f$. For this reason, the interactions in Eq.~\eqref{eq:Fermint} will be neglected in the remainder of this paper, unless otherwise noted. Nevertheless, since they can be important in hadron collider processes \cite{DeSimone:2012fs}, where $p/f \sim m_\ast/f \sim O(1)$ with $m_\ast$ the mass of a resonance, we will return to them in the discussion of the LHC and future collider prospects in Sec.~\ref{Sec:TPsignals}.

Resonances in the gauge sector are assumed to follow the generalized hidden local symmetry approach \cite{Bando:1987br}, where given a $\mathcal{G}/\mathcal{H}$ sigma model, the vector resonances are introduced as gauge fields of a local $\mathcal{G}$ symmetry. In our case $\mathcal{G} = SO(7)$, whose adjoint representation decomposes as $\mathbf{21}=\mathbf{15}\oplus\mathbf{6}$ under $SO(6)$. Thus we introduce vector resonances in the adjoint $\rho_\mu \sim \mathbf{15}$ and in the fundamental $a_\mu \sim \mathbf{6}$ of $SO(6)$. Their Lagrangian is given in App.~\ref{app:A}. 

\section{Scalar potential and realistic EWSB} \label{sec:scalarpotential}
The explicit $SO(7)$ breakings introduced by the weak gauging of $SU(2)_L \times U(1)_Y$ and by the fermionic elementary-composite mixing parameters in Eq.~\eqref{eq:fermion}, which we will often collectively denote by $\epsilon$, generate a radiative potential for the GBs. This can be computed at $1$-loop using the standard Coleman-Weinberg (CW) technique \cite{Coleman:1973jx}. In the unitary gauge and expanded to quartic order in the fields, the effective potential takes the form
\begin{equation} \label{eq:GBPotential}
V(\tilde{h},\chi ) = \frac{1}{2}\mu_h^2 \tilde{h}^2+\frac{\lambda_h}{4}\tilde{h}^4+ \mu_{\rm DM}^2 \chi^* \chi +\lambda_{\rm DM} (\chi^* \chi)^2 + \lambda \tilde{h}^2 \chi^* \chi\, .
\end{equation}
This potential must, first of all, yield a correct EWSB VEV, $\langle \tilde{h} \rangle = v \ll f$. Even though $U(1)_{\rm DM}$ is exactly preserved by the Lagrangian, in general it may still be broken spontaneously. Since this would spoil the DM stability, in the following we only consider parameter choices that satisfy $\left\langle \chi\right\rangle = 0\,$. Then the masses of the physical scalars are
\begin{equation}
m_h^2 = (1-\xi)  \left.\frac{\partial^2 V}{\partial \tilde{h}^2}\right|_{\tilde{h}\,=\,v,\,\chi\,=\,0} = (1-\xi) 2 \lambda_h v^2  \,, \qquad m_\chi^2 = \left.\frac{\partial^2 V}{\partial\chi \partial\chi^*}\right|_{\tilde{h}\,=\,v,\,\chi\,=\,0}=\mu_{\rm DM}^2 +  \lambda v^2\,,
\label{eq:HiggsMass}
\end{equation}
\enlargethispage{-15pt}where the $(1-\xi)$ factor in the expression of $m_h^2$ is due to Eq.~\eqref{eq:canonicalh}. In general, the mass parameters $\mu^2_{h}, \mu^2_{\rm DM}$ and couplings $\lambda_h, \lambda_{\rm DM}, \lambda$ are quadratically and logarithmically sensitive, respectively, to the UV cutoff $\Lambda \lesssim 4\pi f$ of the effective theory.\footnote{Notice that by naive power counting, the quartic couplings can also be quadratically divergent. However, the structure of the field-dependent mass matrices leads to a quadratically divergent term \mbox{$\sim \Lambda^2 \mathrm{STr}\,m^2(h,\chi) = \Lambda^2(k_0 + k_h h^2 + k_\chi \chi^\ast \chi)$} with $k_{0,h,\chi}$ field-independent constants. Thus the leading degree of divergence of the quartics is only logarithmic.} However, to retain predictivity we assume that they are fully saturated by the contribution of the SM fields plus the first few vector and fermion resonances that we introduced in Sec.~\ref{subs:resonances}. This is achieved by imposing a set of generalized WSRs \cite{Weinberg:1967kj}, which ensure that the form factors determining the parameters of the CW potential vanish sufficiently fast at large momenta \cite{Marzocca:2012zn,Pomarol:2012qf}. In addition, we assume that further explicit breakings of $SO(7)$ originating from the UV dynamics, if present, give a subleading contribution to the scalar potential.\footnote{Notice also that, due to the contribution of top quark and SM gauge boson loops, the expression of $\lambda_h$ in Eq.~\eqref{eq:GBPotential} is infrared (IR) divergent. To retain full predictivity, this issue is resolved by adding to $V(\tilde{h},\chi)$ an additional quartic for $\tilde{h}$ that is non-analytic at $\tilde{h}=0$. See App.~\ref{app:B} for further details.}
 
\enlargethispage{-55pt}
Beginning with the gauge sector, we recall that the gauging of $SU(2)_L \times U(1)_Y$ preserves $U(1)_{\rm DM}$ (see e.g. Fig.~\ref{fig:GenBases}), hence the associated loops only yield a contribution to the Higgs mass parameter, denoted $\mu^2_{h,g}$, and one to the quartic coupling, $\lambda_{h,g}$. The UV-finiteness of these coefficients can be obtained by introducing one multiplet of vector resonances in the adjoint of $SO(6)$, $\rho_\mu$, and one in the fundamental, $a_\mu$, and imposing two WSRs that translate into the conditions
\begin{equation} \label{eq:gaugeWSR}
2 f_\rho^2 - 2 f_a^2 = f^2\,,\qquad\quad f_\rho^2 m_\rho^2 = f_a^2 m_a^2 \,, \quad\qquad (\mathrm{WSR}\; 1+2)_{g}
\end{equation}
where $f_{\rho,\,a}$ are the decay constants of the resonances, and $m_{\rho,\,a}$ their masses. The first relation removes the quadratic divergence in $\mu_{h,g}^2$ and makes $\lambda_{h, g}$ finite, whereas the second ensures the cancellation of the residual logarithmic divergence in $\mu_{h,g}^2$. Equations \eqref{eq:gaugeWSR} allow us to express $f_a$ and $m_a$ in terms of $f_{\rho}, m_{\rho}$ and $f$; the first one also requires $f_\rho > f/\sqrt{2}$. The contribution to the Higgs mass parameter reads, at leading order in $g^2/g_\rho^2 \ll 1$ (where $g_\rho = m_\rho /f_\rho$) and neglecting the subleading hypercharge coupling, 
\begin{equation}
\mu_{h,\,{g}}^2 \approx \frac{9\,g^2}{32\pi^2} m_\rho^2\, \frac{f_\rho^2}{f^2} \log\left(\frac{2f_\rho^2/f^2}{2f_\rho^2/f^2 - 1}\right).
\end{equation}
Since this is strictly positive, the gauge loops alone do not lead to EWSB. However, a negative contribution to $\mu_h^2$ can easily arise from the fermionic sector, and $\mu_{h,\,{g}}^2$ will be tuned against it to obtain a realistic Higgs VEV $v \ll f$. On the other hand, the gauge contribution to the Higgs quartic is small, and plays a subleading role.

In the fermionic sector, the elementary-composite mixing parameters $\epsilon$ explicitly break the shift symmetries protecting both $\tilde{h}$ and $\chi$, therefore in general fermion loops yield contributions to all the coefficients in the effective potential of Eq.~\eqref{eq:GBPotential}. To ensure their UV finiteness, we impose two sets of WSRs, which translate into the relations
\begin{align}
\sum_{i=1}^{N_Q} \left| \epsilon_{qQ}^i\right|^2 & = \sum_{j=1}^{N_S} \left|\epsilon_{qS}^j \right|^2,\qquad\quad
&\sum_{i=1}^{N_Q} \left| \epsilon_{tQ}^i \right|^2 & = \sum_{j=1}^{N_S} \left| \epsilon_{tS}^j \right|^2\,, & (\mathrm{WSR}\; 1)_f \label{eq:WSR1}\\
\sum_{i=1}^{N_Q} \left| \epsilon_{qQ}^i \right|^2 m_{Q_i}^2 & = \sum_{j=1}^{N_S} \left| \epsilon_{qS}^j \right|^2 m_{S_j}^2\,,
&\sum_{i=1}^{N_Q} \left| \epsilon_{tQ}^i \right|^2 m_{Q_i}^2 & = \sum_{j=1}^{N_S} \left| \epsilon_{tS}^j \right|^2 m_{S_j}^2\,. & (\mathrm{WSR}\; 2)_f \label{eq:WSR2}
\end{align}
The first set of WSRs reduce the $1$-loop degree of divergence of the mass parameters $\mu^2_{h,f}$ and $\mu^2_{\mathrm{DM},f}$ (where the ``$f$'' subscript indicates the fermionic piece) from quadratic to logarithmic and make the dimensionless couplings finite, whereas the second set remove the residual logarithmic divergences in $\mu^2_{h,f}$ and $\mu^2_{\mathrm{DM},f}$. The minimal set of resonances compatible with Eqs.~(\ref{eq:WSR1}, \ref{eq:WSR2}) consists of one $SO(6)$ fundamental $Q$ and one singlet $S$. This `one-layer' setup is very predictive, but, as discussed in Sec.~\ref{sec:1layer}, it leads to a DM candidate that is phenomenologically ruled out. Nevertheless, thanks to the simplicity of the one-layer model, we obtain some analytical results and thus gain valuable insight. We then turn to an enlarged setup where two copies of each species of resonance are present below the cutoff. As shown in Sec.~\ref{sec:2layers}, this ``two-layer'' construction gives sufficient freedom to accommodate a fully viable DM candidate, leading us to concentrate on this model for our phenomenological analysis.

\subsection{One layer of fermionic resonances} \label{sec:1layer}
We consider the fermionic Lagrangian of Eq.~\eqref{eq:fermion} with $N_Q = N_S = 1$. In this case the WSRs in Eqs.~(\ref{eq:WSR1}, \ref{eq:WSR2}) give
\begin{equation} \label{eq:WSR1layer}
\epsilon_{qQ}^2 = \epsilon_{qS}^2\,,\qquad \epsilon_{tQ}^2 = \epsilon_{tS}^2\,,\qquad  m^2_Q = m^2_S\,,
\end{equation}
where we have assumed all the parameters to be real, so that $CP$ is conserved. In the following we take, without loss of generality, positive masses $m_Q = m_S \equiv m > 0$. Then the conditions in Eq.~\eqref{eq:WSR1layer} do not fix the relative signs of the mixing parameters, $\epsilon_{qQ} =\pm\, \epsilon_{qS}\,$ and $\epsilon_{tQ} =\pm\, \epsilon_{tS}\,$. If these two signs are equal, then the non-derivative part of Eq.~\eqref{eq:fermion} has an additional $SO(7)$ symmetry that allows the Goldstone matrix to be removed by means of a field redefinition (see for example Ref.~\cite{Grojean:2013qca}), hence the scalar potential vanishes. If instead the mixings have opposite sign, the potential does not vanish. Taking for definiteness $\epsilon_{qQ} = - \epsilon_{qS} \equiv - \epsilon_q\,$ and $\epsilon_{tQ} = \epsilon_{tS} \equiv \epsilon_t\,$,\footnote{Notice that by redefining the phases of the resonances, we can equivalently choose a field basis with same-sign mixings and $m_Q + m_S = 0$. This is a realization of the ``maximal symmetry'' of Ref.~\cite{Csaki:2017cep}. Accordingly, the tuning of the model is minimal, see Eq.~\eqref{eq:1layertuning} below.} we find
\begin{equation} \label{eq:1layerDM}
\mu_{\mathrm{DM},f}^2 = \lambda_{\mathrm{DM},f} = 0\,,\qquad \lambda_f = - \frac{\mu_{h,f}^2}{f^2} = \frac{N_c \epsilon_q^2 \epsilon_t^2 m^2 \log (M_T^2/M_S^2)}{2\pi^2 f^4 (M_T^2 - M_S^2)}\,,
\end{equation}
where $M_{T,S}^2 = m^2 + \epsilon_{q,t}^2$ are the squared masses of the top partners that mix with the $q_{L}$ and $t_R$, respectively, neglecting small corrections due to EWSB. Equation \eqref{eq:1layerDM} gives the complete expressions of $\mu_{\rm DM}^2, \lambda_{\rm DM}$ and $\lambda$, which do not receive any contribution from the gauge sector. In addition, we find the following approximate expression for the Higgs quartic,
\begin{equation} \label{eq:1layerh}
\lambda_h \approx \frac{N_c \epsilon_q^2 \epsilon_t^2 m^2 \log (M_T^2/M_S^2)}{\pi^2 f^4 (M_T^2 - M_S^2)}\,,
\end{equation}
obtained by neglecting the gauge contribution to the potential. Equations \eqref{eq:1layerDM} and \eqref{eq:1layerh} suggest the relation $\lambda \approx \lambda_h/2$, which is indeed verified within $20\%$ in our numerical scan of the parameter space. Therefore both the portal coupling and the DM mass are fixed in terms of $v$ and the Higgs mass, 
\begin{equation} \label{eq:DM1layer}
\lambda \approx \frac{\lambda_h}{2} \simeq \frac{m_h^2}{4v^2} \simeq 0.065\,, \qquad m_{\chi}^2 = \lambda v^2 \approx \frac{\lambda_h v^2}{2} \simeq \frac{m_h^2}{4} \ \simeq (63\;\mathrm{GeV})^2\,. 
\end{equation}
Unfortunately, this combination of DM mass and coupling has already been ruled out experimentally: since the DM is light and the portal coupling is not very suppressed, the derivative interactions in Eq.~\eqref{eq:GBLagrangian} have negligible effects, and the phenomenology of $\chi$ can be approximately described with a renormalizable Higgs portal model \cite{Silveira:1985rk,McDonald:1993ex,Burgess:2000yq}. In this model, the region $\lambda \sim \lambda_h/2$, $m_\chi \sim m_h/2$ has been ruled out by direct detection experiments and, for $m_\chi < m_h/2$, also by LHC bounds on the Higgs invisible width, see e.g. Ref.~\cite{Casas:2017jjg} for a recent assessment. 

The problematic values in Eq.~\eqref{eq:DM1layer} arose because in the presence of only one layer of resonances, the second set of WSRs in Eq.~\eqref{eq:WSR2} imply that the form factors $\Pi_{L_1}$ and $\Pi_{R_1}$ in Eqs.~(\ref{eq:fermiFFs0},~\ref{eq:fermiFFs}) vanish, and as a consequence we find a non-generic form of the potential, whose structure is entirely determined by the top mass form factor $\Pi_{LR}$. Thus it seems plausible that a viable phenomenological scenario may be obtained by extending the model to include a second layer of resonances, which provides additional parametric freedom and should allow for significant departures from Eq.~\eqref{eq:DM1layer} while preserving full calculability via WSRs. This hypothesis is supported by a test on the one-layer model, where we lift the second set of WSRs and instead cut off the residual logarithmic divergences in $\mu_{h,f}^2$ and $\mu_{\mathrm{DM},f}^2$ at the scale $\Lambda = 4\pi f$. In this case the potential has a generic form, and accordingly we find that large deviations from Eq.~\eqref{eq:DM1layer} are realized. Therefore, in the next subsection we will analyze the model where two layers of fermionic resonances lie below the cutoff. Before doing so, however, we point out a few additional properties of the case $N_Q = N_S = 1$, which apply at least at the qualitative level also in the extended model. Combining Eq.~\eqref{eq:1layerh} with the expression of the top mass at leading order in $\xi \ll 1$, $m_t \simeq \sqrt{2} \epsilon_q \epsilon_t v /(M_T M_S f)$, we obtain
\begin{equation} \label{eq:hTP}
\frac{m_h^2}{m_t^2} \approx \frac{N_c}{\pi^2 f^2} \frac{M_T^2 M_S^2}{M_T^2 - M_S^2} \log (M_T^2/M_S^2)\,.
\end{equation}
This relation, which was already obtained in the context of the MCHM based on $SO(5)/SO(4)$ \cite{Matsedonskyi:2012ym,Pomarol:2012qf}, shows that realizing a light Higgs requires at least one of the top partners to be relatively light, with mass roughly comparable to $f$. Equation \eqref{eq:hTP} is verified numerically to good accuracy, with minor corrections arising due to the presence of the gauge contribution in the potential, which was neglected in the derivation of Eq.~\eqref{eq:1layerh}. The fine-tuning needed to obtain $v \ll f$ can be estimated using the standard measure \cite{Barbieri:1987fn}
\begin{equation} \label{eq:tuningGeneral}
\Delta = \Delta_\xi = \text{max}_i \left|\frac{\partial \log\xi}{\partial \log c_i}\right|\,,
\end{equation}
where $c_i$ denotes the input parameters. In the one-layer model we have $c_i = \lbrace \epsilon_q ,\epsilon_t, m, f_\rho, m_\rho\rbrace$, but an immediate estimate of the tuning can be obtained by noticing that if the gauge contribution to $V$ is neglected, Eqs.~(\ref{eq:1layerDM})-(\ref{eq:DM1layer}) give $\xi \approx 1/2$. Thus $\mu_{h,g}^2$ must be adjusted to give $\xi \ll 1$, leading to a fine-tuning
\begin{equation} \label{eq:1layertuning}
\Delta^{-1} \sim 2\xi \,.
\end{equation}
This is in fact the minimal (or irreducible) amount of tuning characteristic of models where the Higgs potential is entirely generated at the radiative level. A numerical estimate obtained using Eq.~\eqref{eq:tuningGeneral} agrees well with this result.

To conclude, we remark that very similar results, including the prediction of Eq.~\eqref{eq:DM1layer}, were previously found in Ref.~\cite{Marzocca:2014msa} for the realization of the $SO(6)/SO(5)$ model with minimal fermion content. 

\subsection{Two layers of fermionic resonances} \label{sec:2layers}
We consider the fermionic Lagrangian of Eq.~\eqref{eq:fermion} with $N_Q = N_S = 2$. In this case the conditions imposed by the first set of WSRs, Eq.~\eqref{eq:WSR1}, can be solved in terms of two mixings $\epsilon_{q,t}$ and four angles $\alpha , \theta, \beta$ and $\phi\,$,
\begin{equation}
\frac{\epsilon_{qQ}^1}{\cos\alpha} = \frac{\epsilon_{qQ}^2}{\sin\alpha}  =  \frac{\epsilon_{qS}^1}{\cos\theta} = \frac{\epsilon_{qS}^2}{\sin\theta} = \epsilon_q\,, \qquad 
\frac{\epsilon_{tQ}^1}{\cos\beta} = \frac{\epsilon_{tQ}^2}{\sin\beta}  =  \frac{\epsilon_{tS}^1}{\cos\phi} = \frac{\epsilon_{tS}^2}{\sin\phi} = \epsilon_t\,.
\end{equation}
The second set of WSRs in Eq.~\eqref{eq:WSR2} fixes two of the angles, modulo discrete ambiguities. We choose
\begin{equation} \label{eq:angles2layers}
s^2_{\theta,\phi} = \frac{m_{Q_1}^2-m_{S_1}^2+\left(m_{Q_2}^2-m_{Q_1}^2\right) s^2_{\alpha,\beta}}{m_{S_2}^2-m_{S_1}^2}\,\qquad (s^2_{x} \equiv \sin^2 x),
\end{equation}
and without loss of generality we assume $m_{S_2} > m_{S_1}$ and $m_{Q_2} > m_{Q_1}$. The resulting parameter space\footnote{Note that for special values of the parameters, the model can be realized via a three-site construction \cite{Panico:2011pw}.} is scanned numerically, see App.~\ref{app:B} for details.  
\begin{figure}[t]
\centering
\includegraphics[width=0.495\textwidth]{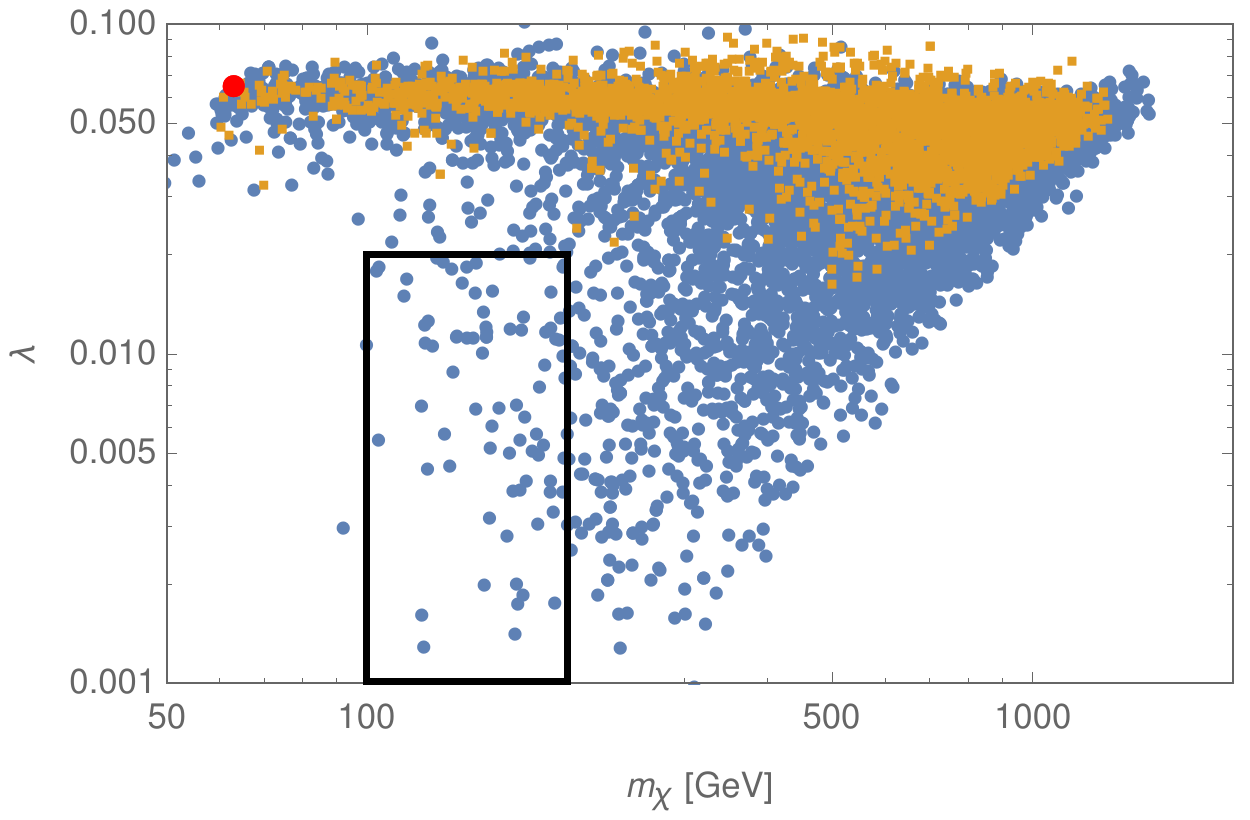}
\includegraphics[width=0.495\textwidth]{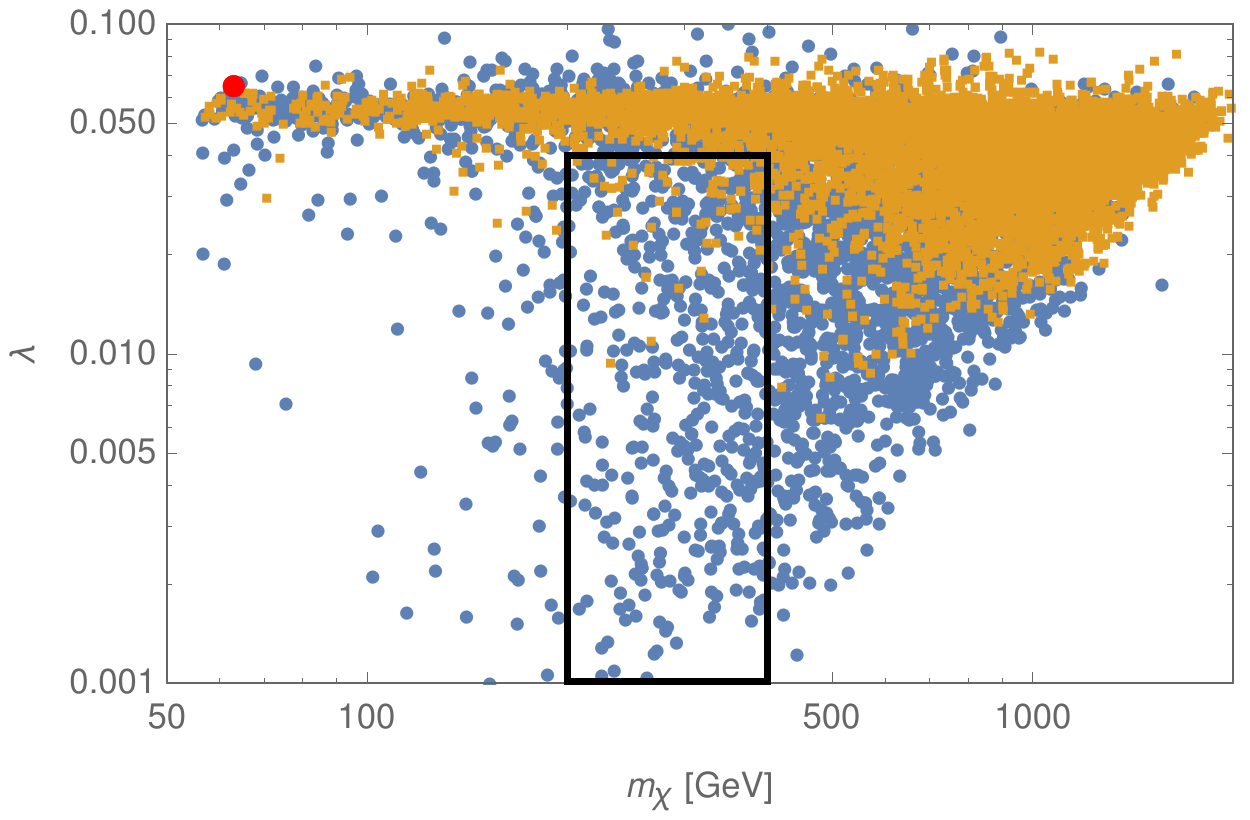}
\caption{Distribution in the $(m_\chi, \lambda)$ plane for the parameter scan of the two-layer model. The left panel corresponds to $f = 1$ TeV, the right panel to $f = 1.4$ TeV. The black boxes roughly indicate the viable regions of parameters for DM. The red dot shows the approximate prediction of the one-layer model, Eq.~\eqref{eq:DM1layer}. For orange (blue) points, the lightest fermionic resonance is heavier (lighter) than the approximate LHC lower bound of $1$ TeV.}
\label{fig:2layerscan}
\end{figure}
Figure \ref{fig:2layerscan} shows the resulting distribution in the $(m_\chi, \lambda)$ plane for two values of $f$, namely $1$ TeV and $1.4$ TeV. As expected, large deviations from the predictions of $1$-loop-finite one-layer model are generic. First of all, $\chi$ is typically much heavier than $m_h/2 \sim 63$ GeV. In particular, its mass populates the $100\,$-$\,400$ GeV range where, as will be shown in Sec.~\ref{sec:pheno}, we find that the DM relic abundance is around the observed value. In addition, the portal coupling $\lambda$ can be smaller than $\lambda_h/2 \sim 0.065$. This is crucial because, as will also be discussed in detail in Sec.~\ref{sec:pheno}, direct detection bounds require smaller values of this coupling. In Fig.~\ref{fig:2layerscan} we also observe that a reduction of the portal coupling is correlated with the appearance of light top partners, which can run into tension with the current lower bound of approximately $1$ TeV set by LHC searches. (We will discuss the LHC constraints in detail in Sec.~\ref{sec:LHC}, but this rough estimate suffices for the scope of the present discussion.) In fact, for $f = 1$ TeV we do not find any points that have viable DM parameters, i.e. roughly $100\;\mathrm{GeV} \lesssim m_\chi \lesssim 200\;\mathrm{GeV}$ and $\lambda \lesssim 0.02$ (indicated by the black box in the left panel of Fig.~\ref{fig:2layerscan}), without running into conflict with top partner bounds. Increasing $f$ relaxes this tension, because it allows the top partners to be naturally heavier and it shifts the viable DM mass region to higher values, where the constraints on $\lambda$ from direct detection are less stringent. The minimal $f$ that yields a sizable region of allowed parameter space is $1.4$ TeV, which we will therefore use as our primary benchmark for the remainder of this paper. The corresponding viable ranges for the DM mass and portal coupling are $200\;\mathrm{GeV} \lesssim m_\chi \lesssim 400\;\mathrm{GeV}$ and $\lambda \lesssim 0.04$, respectively, shown by the black box in the right panel of Fig.~\ref{fig:2layerscan}.
\begin{figure}[t]
\centering
\includegraphics[width=0.485\textwidth]{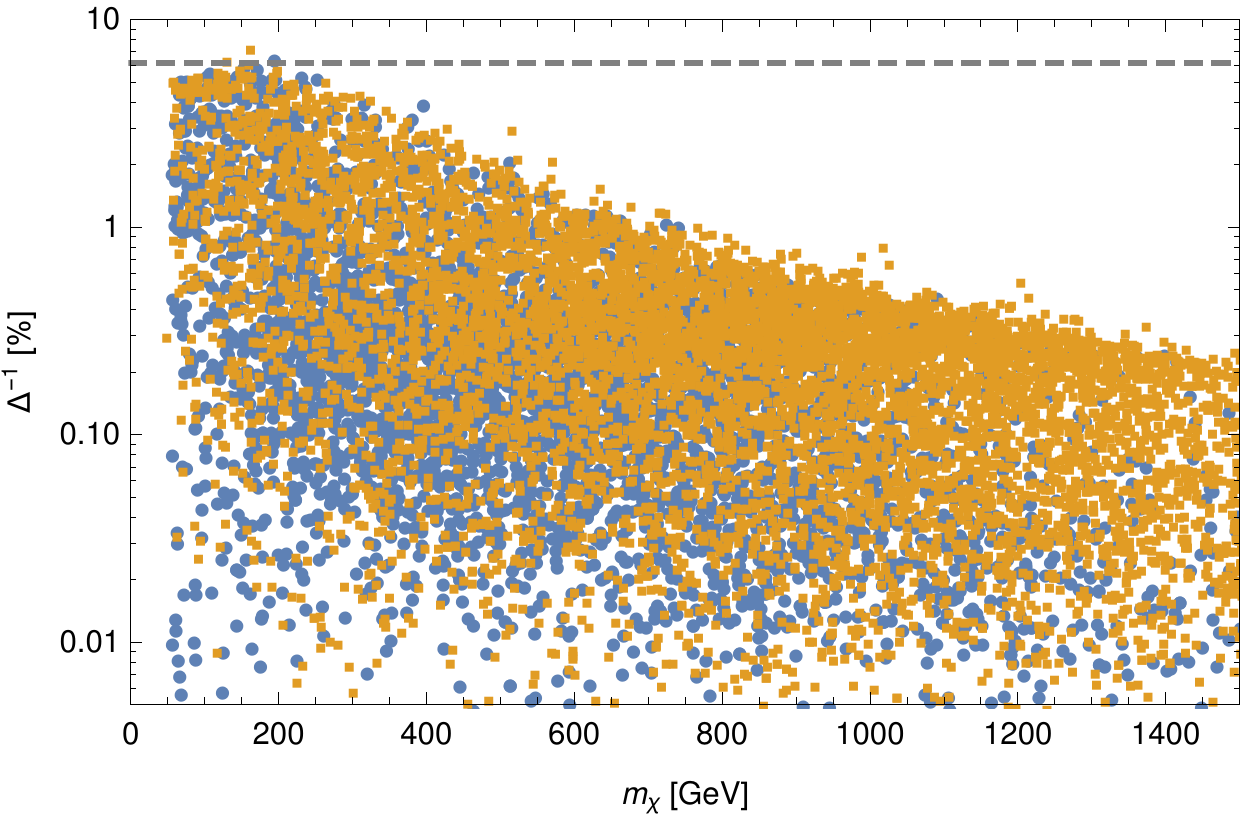}
\includegraphics[width=0.495\textwidth]{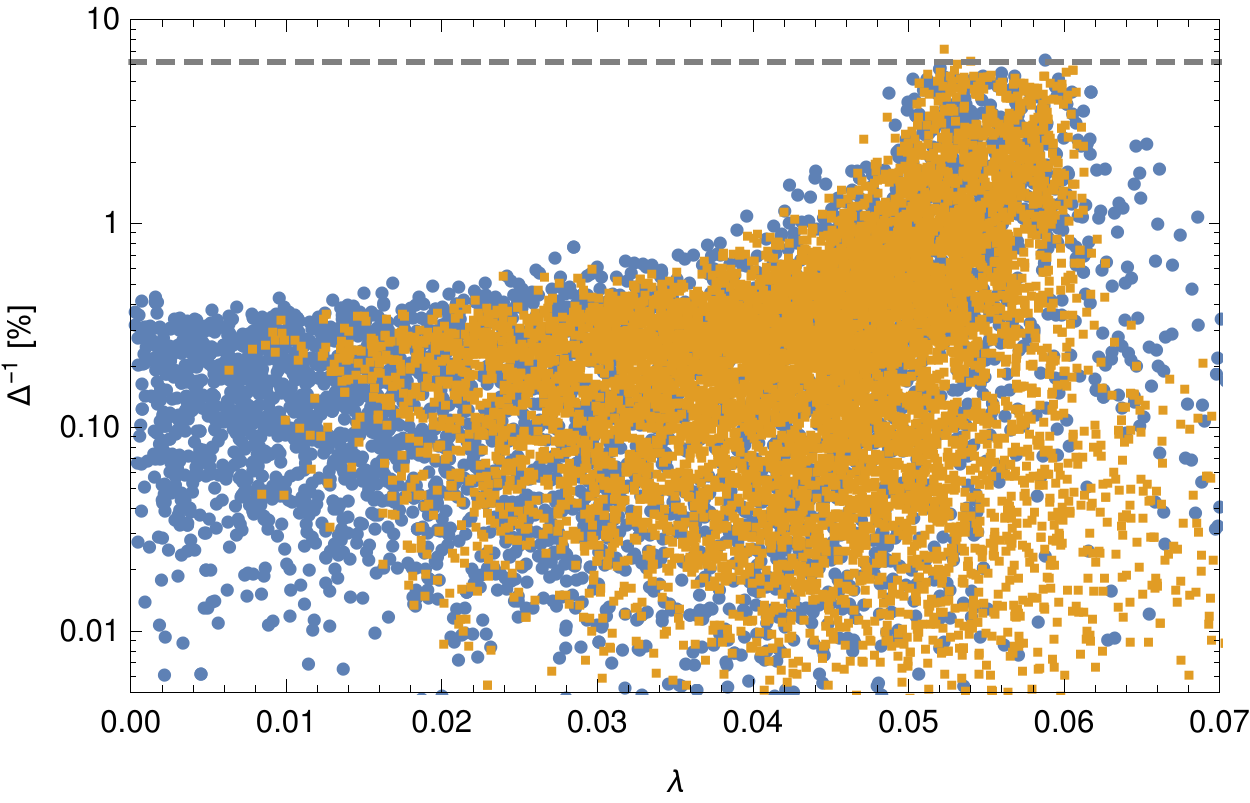}
\caption{Fine-tuning of the two-layer model, shown versus the DM mass (left panel) and versus the portal coupling (right panel). For orange (blue) points, the lightest fermionic resonance is heavier (lighter) than the approximate LHC lower bound of $1$ TeV. The scale $f$ is fixed to $1.4$ TeV.}
\label{fig:2layertuning}
\end{figure}

The irreducible tuning associated to $f = 1.4$ TeV is, according to Eq.~\eqref{eq:1layertuning}, $\Delta^{-1} \sim 2 \xi \simeq 6 \%$. A more precise, point-by-point estimate is obtained by applying the general definition of Eq.~\eqref{eq:tuningGeneral}, and shown in Fig.~\ref{fig:2layertuning}. We see that as the departure from the predictions of the one-layer model becomes larger, namely as the $\chi$ mass is raised to $m_\chi \gg m_h/2$ and the portal coupling is suppressed to $\lambda \ll \lambda_h/2\,$, the minimum tuning required increases. The worsening of the tuning for larger $m_\chi$, observed in the left panel of Fig.~\ref{fig:2layertuning}, can be explained by noticing that a heavier $\chi$ can only be obtained by increasing the size of the form factor $\Pi_{R_1}$, which vanishes in the one-layer model (see Eq.~\eqref{eq:FFs}). This in turn requires a more severe cancellation in the Higgs mass parameter in order to achieve a small $\xi$. Nevertheless, a phenomenologically viable DM mass, $200\;\mathrm{GeV} \lesssim m_\chi \lesssim 400\;\mathrm{GeV}$, can be obtained without significantly exacerbating  the tuning compared to irreducible contribution of $2 \xi \sim 6\%$. On the other hand, from the right panel of Fig.~\ref{fig:2layertuning} we read that a portal coupling that is small enough to satisfy the current direct detection bounds, $\lambda \lesssim 0.04$, requires $\Delta^{-1} \lesssim 1\%$. We have also checked that once the Higgs VEV and mass are fixed to the observed values, no additional tuning is needed in the DM mass: replacing $\xi$ with $\mu_{\rm DM}^2$ in Eq.~\eqref{eq:tuningGeneral}, for the points shown in Fig.~\ref{fig:2layertuning} we found that $\Delta^{-1}_{\mu^2_{\rm DM}}$ can be of $O(1)$ even for DM mass as low as $200$ GeV. In summary, we estimate that in this model the level of fine-tuning required to solve both the Higgs naturalness and DM puzzles is $1\%$ or slightly worse. This is primarily a consequence of the experimental pressure from direct detection experiments and LHC direct searches for top partners. 
\begin{figure}[t]
\centering
\includegraphics[width=0.485\textwidth]{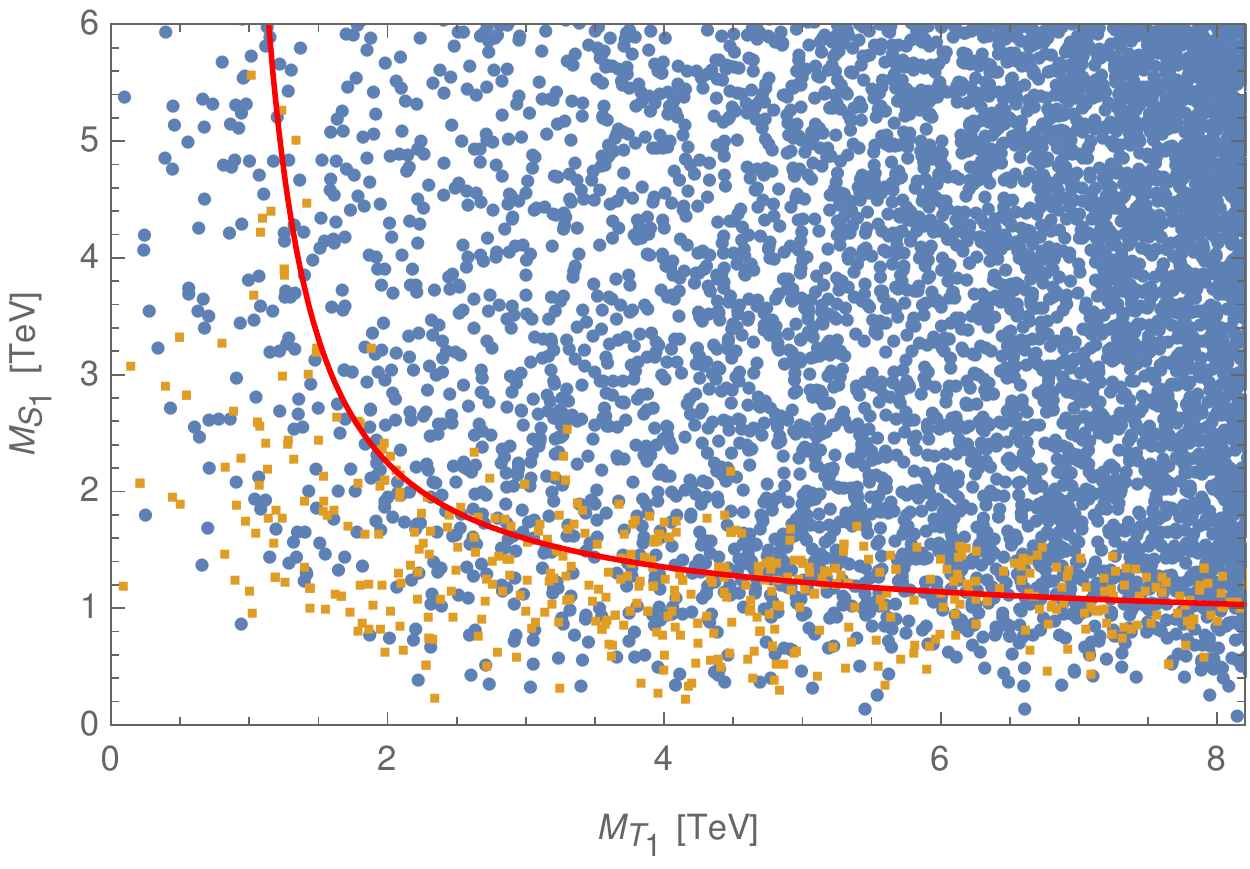}
\caption{Mass of the lightest top partner mixing with the $t_R$ ($M_{S_1}$) versus mass of the lightest top partner mixing with the $q_L$ ($M_{T_1}$), neglecting EWSB corrections, in the two-layer model. Orange (blue) points have a Higgs mass within (outside) the range $120\;\mathrm{GeV} < m_h < 130$ GeV. The red line shows the approximate prediction of the one-layer model, Eq.~\eqref{eq:hTP}. We set $f = 1.4$ TeV.}
\label{fig:2layerhmass}
\end{figure}

Figure \ref{fig:2layerhmass} shows that the correlation between a light Higgs and light top partners, which in the one-layer model was expressed by Eq.~\eqref{eq:hTP}, holds in the two-layer setup as well. Furthermore, Eq.~\eqref{eq:hTP} still yields a reasonable quantitative first approximation, provided we identify $M_T$ and $M_S$ with the masses of the lightest top partners mixing with $q_L$ and $t_R$, respectively. 

Having qualitatively characterized the viable parameter space, we are now ready to present its phenomenology. We begin in Sec.~\ref{sec:pheno} with DM physics, and then discuss the collider aspects in Sec.~\ref{sec:LHC}.

\section{Dark matter phenomenology} \label{sec:pheno}
In this section we present the phenomenology of our DM candidate $\chi$. We focus on  two main observables, namely the DM relic abundance and the DM-nucleus scattering cross section, which is relevant for direct detection experiments. We conclude the section with a brief comment on the constraints from indirect detection. 
\subsection{Effective theory for DM annihilation} \label{sec:annEFT}
The DM relic abundance is set by the annihilation rate in the early universe, which takes place at an energy scale $\sqrt{s} \sim 2 m_\chi \ll m_\ast$, where $m_\ast$ denotes the mass of the strong sector resonances ($m_\ast \sim g_\ast f$, with $g_\ast$ some strong sector coupling). The relic abundance can therefore be calculated in an effective theory where the resonances have been integrated out, and only the pNGB scalars $\chi, h$, the SM gauge bosons and the SM fermions are included as propagating degrees of freedom. Assuming that the freeze-out temperature satisfies $T_f \ll v$, which is generically the case for DM with a weak-scale mass, the Lagrangian can be written in the broken electroweak phase. Additionally, we will consider operators which are at most quadratic in the DM field, since higher-order terms do not contribute to the annihilation processes. The effective Lagrangian has the structure
\begin{equation} \label{eq:Leff_ann}
\mathcal{L}_{\rm eff}\; =\; \underbrace{\mathcal{L}_{\rm GB}\; +\; \mathcal{L}_t}_{\text{tree}} \; \underbrace{-\; V_{\rm eff}}_{\text{1-loop}}\,.
\end{equation}
The first piece originates from the sigma model Lagrangian in Eq.~\eqref{eq:GBLagrangian}, expanded in terms of the physical fields
\begin{align}
\mathcal{L}_{\rm GB} &\,=\, \frac{1}{2}(\partial_\mu h)^2 \Big( 1 + 2\, a_{hhh}\frac{h}{v}+ 2\, a_{hh\chi\chi}\frac{\chi^* \chi}{v^2} \Big) + \partial_\mu \chi \partial^\mu \chi^\ast + \frac{1}{v} \partial_\mu h\,\partial^\mu (\chi^* \chi)\Big(b_{h\chi\chi} + b_{hh\chi\chi}\frac{h}{v}\Big) \nonumber \\
 \quad &\, + 2\, a_{hVV} \frac{h}{v}\Big(m_W^2 W_\mu^+ W^{-\,\mu}  + \frac{m_Z^2}{2} Z_\mu Z^\mu\Big). \label{eq:LEGBint}
\end{align}
$V_{\rm eff}$ arises instead from the radiative scalar potential, Eq.~\eqref{eq:GBPotential}, and reads
\begin{equation} \label{eq:Veff}
V_{\rm eff} = \frac{1}{2}m_h^2 h^2 + d_{hhh}\,\frac{m_h^2}{2v}\, h^3 + m_\chi^2 \chi^*\chi + 2\,d_{h\chi\chi} v\lambda  h\chi^*\chi + d_{hh\chi\chi}\lambda h^2 \chi^*\chi\,.
\end{equation}
The scalar couplings in Eq.~\eqref{eq:Veff}, despite being loop-suppressed, can have effects comparable to those of the tree-level interactions in $\mathcal{L}_{\rm GB}$, whose derivative structure leads to a suppression $\sim s/f^2 \ll 1\,$ (see Eq.~\eqref{eq:cancellation} below) \cite{Frigerio:2012uc}. With the exception of $\lambda$, all the dimensionless coefficients in Eqs.~(\ref{eq:LEGBint},~\ref{eq:Veff}) are functions of $\xi$ only and are given in Eq.~\eqref{eq:EFTcouplingsANN}.

\begin{figure}[t]
\centering
\includegraphics[width=0.485\textwidth]{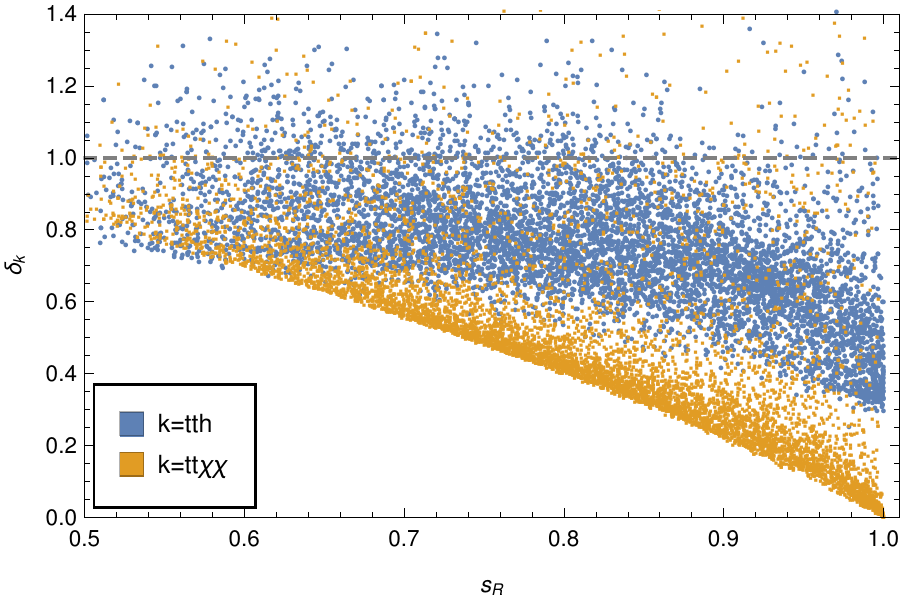}
\caption{Corrections from top partner mixing to the effective $t\bar{t}h$ and $t \bar{t} \chi^\ast \chi$ couplings, defined as $\delta_{tt\chi\chi} \equiv c_{tt\chi\chi}/c_{tt\chi\chi}^{\rm nl\sigma m}$ and $\delta_{tth} \equiv (c_{tth} - 1)/(c_{tth}^{\rm nl\sigma m} - 1)$, as functions of the compositeness fraction $s_R$ of the right handed top (see Eq.~\eqref{eq:tRcmpfraction} for its definition). The gray dashed line indicates the pure sigma model result, where top partner mixing is neglected. The points shown are obtained from a parameter scan of the two-layer model with $f=1.4$ TeV, requiring all fermionic resonances to be heavier than the approximate LHC bound of $1$ TeV.}
\label{fig:TP_coupl_Corr}
\end{figure}

Finally, the Lagrangian containing the couplings of the top quark relevant to DM annihilation is
\begin{equation}
\mathcal{L}_{t} = i\bar{t}\slashed{\partial} t - m_t \bar{t} t\Big( 1 + c_{tth}\frac{h}{v} + 2\, c_{tt\chi\chi}\frac{\chi^\ast \chi}{v^2}\Big),
\label{eq:topEffLagr}
\end{equation}
where the dimensionless coefficients have the form
\begin{equation} \label{eq:topcouplings}
c_{k} = c_{k}^{\rm nl\sigma m}(\xi) + O\Big(\xi \frac{\epsilon^2}{m_\ast^2} \Big), \qquad k = \{tth, tt\chi\chi\}.
\end{equation}
The functions $c_{k}^{\rm nl\sigma m}(\xi)$ encode the nonlinearity of the sigma model and read
\begin{equation} \label{eq:topnlsm}
c_{tth}^{\rm nl\sigma m} = \frac{1 - 2 \xi}{\sqrt{1-\xi}}\,,\qquad  c_{tt\chi\chi}^{\rm nl\sigma m} = -  \frac{\xi}{2(1 - \xi)}\,.
\end{equation}
The additional terms in the RHS of Eq.~\eqref{eq:topcouplings} come instead from the mixing of the top with the top partners. These terms are suppressed unless one of the chiralities of the top is largely composite, in which case $\epsilon \sim m_\ast\,$, and were neglected in the previous studies of composite pNGB DM of Refs.~\cite{Frigerio:2012uc,Marzocca:2014msa}. In our analysis, however, we find that these corrections play a very important role, as can be seen in Fig.~\ref{fig:TP_coupl_Corr}, where the full numerical value of the $c_k$ coefficients is compared to the $c_k^{\rm nl\sigma m}(\xi)$.
In particular, the coefficient $c_{tt\chi \chi}$ is strongly suppressed by top partner mixing even for moderate $t_R$ compositeness, and in the limit of fully composite $t_R$ the top partner contribution exactly cancels $c_{tt\chi\chi}^{\rm nl\sigma m}(\xi)$, leading to a vanishing $c_{tt\chi\chi}\,$. This can be understood as follows: With our choice of embeddings, the shift symmetry of the DM pNGB $\chi$ is automatically preserved by the couplings of the elementary $q_L$ to the strong sector resonances, whereas the couplings of the elementary $t_R$ break it (see Eq.~\eqref{eq:embed}). However, in the limit where the physical RH top is a fully composite field (whose overlap with the elementary fermion is zero), its couplings also preserve the $\chi$ shift symmetry, hence a non-derivative $t \bar{t} \chi^\ast \chi$ coupling is forbidden. On the other hand, the $t \bar{t} h$ coupling receives smaller, but still important, corrections from top partner mixing.\footnote{Notice that the $t \bar{t} h$ coupling does not vanish at full RH top compositeness, because even in that limit the coupling of $q_L$ to the strong sector breaks the $h$ shift symmetry.}

\subsection{DM relic abundance} \label{subsec:DMra}
The present abundance of DM, which arises from its freeze-out in the early Universe, is computed by solving the corresponding Boltzmann equation. A useful approximate solution is given by
\begin{equation} \label{eq:DMrelic}
\frac{\Omega_{\rm DM}h^2}{0.1198} \simeq \frac{3 \cdot 10^{-26} \mathrm{cm}^{3}\, \mathrm{s}^{-1}}{\tfrac{1}{2} \left\langle \sigma v_{\rm rel} \right\rangle(T_f)}\,.
\end{equation}
On the LHS of this equation, $\Omega_{\rm DM}$ is the ratio between the energy density of DM and the critical energy density of the Universe, $h = H_0/(100 \;\mathrm{km}/\mathrm{s}/\mathrm{Mpc})$ is the reduced value of the present Hubble parameter, and $(\Omega_{\rm DM} h^2)_{\rm exp} = 0.1198 \pm 0.0015$ is the experimental value as measured by the Planck collaboration \cite{Ade:2015xua}. On the RHS, $\left\langle \sigma v_{\rm rel}\right\rangle (T_f)$ is the thermally averaged annihilation cross section times the relative velocity of two DM particles, computed at the freeze-out temperature $T_f \approx m_\chi/20\,$. The factor $1/2$ in the denominator of the RHS accounts for the fact that the DM is not self-conjugate.

DM annihilation proceeds dominantly via $\chi \chi^\ast \to t\bar{t}, WW, ZZ$ and $hh$. All these processes are mediated by diagrams where a Higgs is exchanged in the $s$-channel. Even though the $\chi \chi^\ast \to hh,\, t\bar{t}$ amplitudes receive additional contributions, it is nevertheless useful to assume in first approximation that annihilation proceeds entirely through $s$-channel Higgs exchange. In this case the cross section is proportional to the square of the $\chi\chi^\ast h$ vertex, which from the effective Lagrangian of Eq.~\eqref{eq:Leff_ann} reads
\begin{figure}[t]
\includegraphics[width=0.495\textwidth]{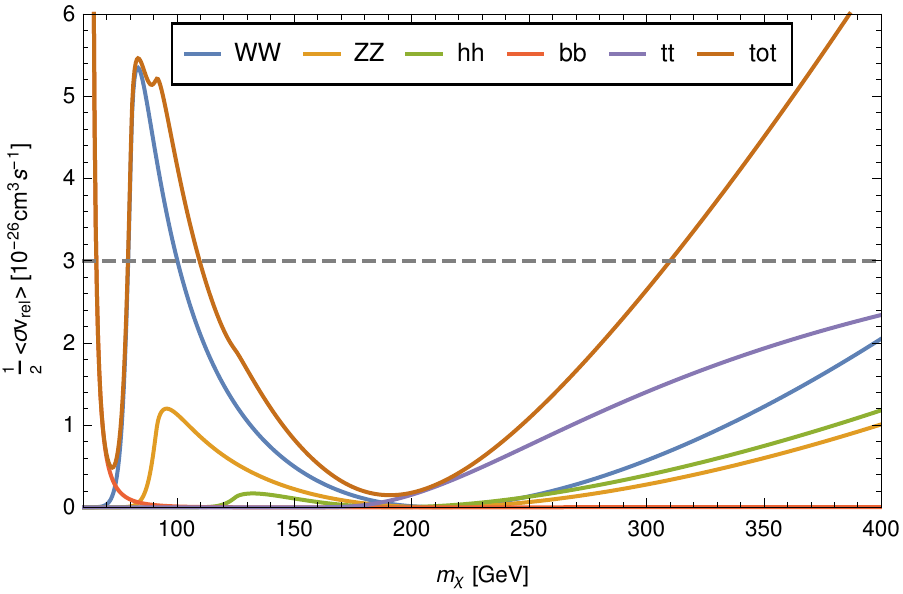}
\includegraphics[width=0.495\textwidth]{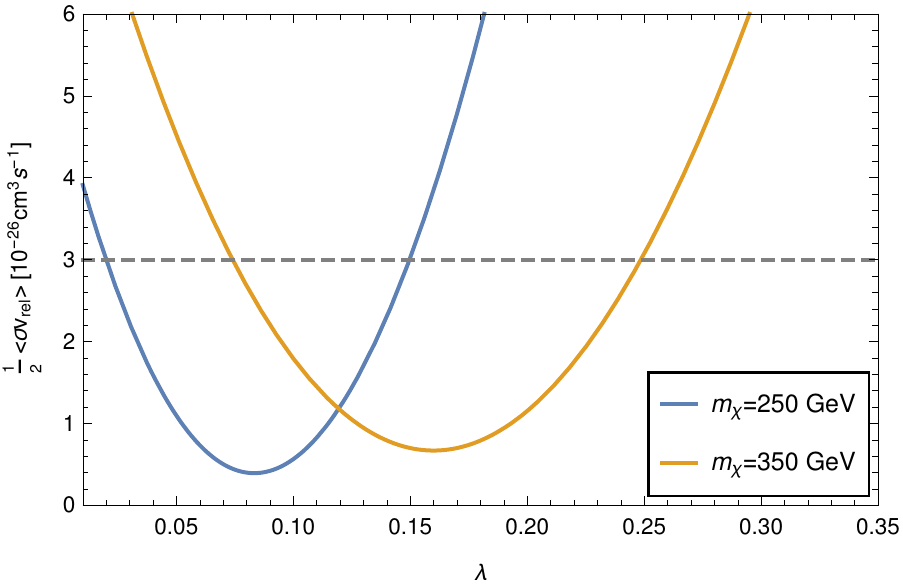}
\caption{Thermally averaged cross section for DM annihilation. The gray dashed line shows the value required to reproduce the present relic abundance according to the approximate relation in Eq.~\eqref{eq:DMrelic}. The scale $f$ was fixed to $1.4$ TeV. In the left panel we set the portal coupling to the representative value $\lambda = 0.05$, in the right panel we chose two representative values of the DM mass. In both panels the $t\bar{t}h$ and $t\bar{t}\chi^\ast \chi$ couplings were set to their sigma model values (Eq.~\eqref{eq:topnlsm}), thus neglecting top partner mixing. With this simplification, $\langle \sigma v_{\rm rel} \rangle$ is completely determined by $f, m_\chi$ and $\lambda$.}
\label{fig:thermXsect}
\end{figure}
\begin{equation} \label{eq:cancellation}
\sigma v_{\rm rel}  \propto \Big(\frac{b_{h\chi\chi}}{v}s - 2\,d_{h\chi\chi} \lambda v \Big)^2 \approx  v^2 \Big(\frac{s}{f^2} - 2 \lambda\Big)^2,
\end{equation}      
where the first term comes from the derivative interactions in Eq.~\eqref{eq:LEGBint} and the second term from the radiative scalar potential in Eq.~\eqref{eq:Veff}. Neglecting relativistic corrections we have $s \approx 4 m_{\chi}^2$, therefore the two contributions cancel out for $m_{\chi}^2 \sim \lambda f^2 /2$, leading to a strong suppression of the annihilation cross section \cite{Frigerio:2012uc,Marzocca:2014msa,Fonseca:2015gva}. This feature can be clearly observed in the cross sections for annihilation into $WW,ZZ$ and also $hh$, see the left panel of Fig.~\ref{fig:thermXsect}. The structure in Eq.~\eqref{eq:cancellation} also implies that for given $f$ and $m_{\chi}$, there are two values of the portal coupling $\lambda$ that reproduce the observed DM relic density, see the right panel of Fig.~\ref{fig:thermXsect}. As will be shown below, however, the branch with larger $\lambda$ is excluded by direct detection, whereas the one with smaller portal coupling provides a viable scenario.

\begin{figure}[t]
\includegraphics[width=0.495\textwidth]{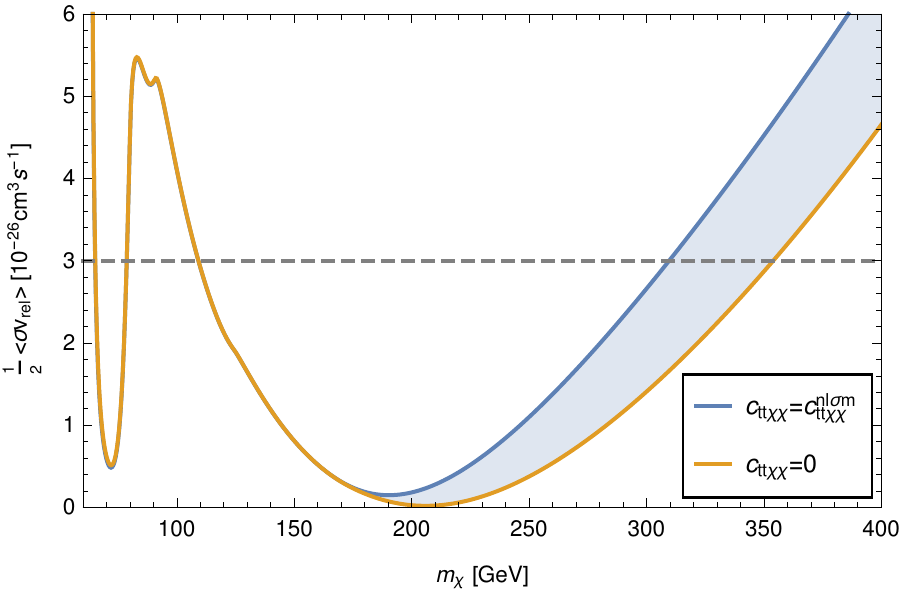}
\includegraphics[width=0.495\textwidth]{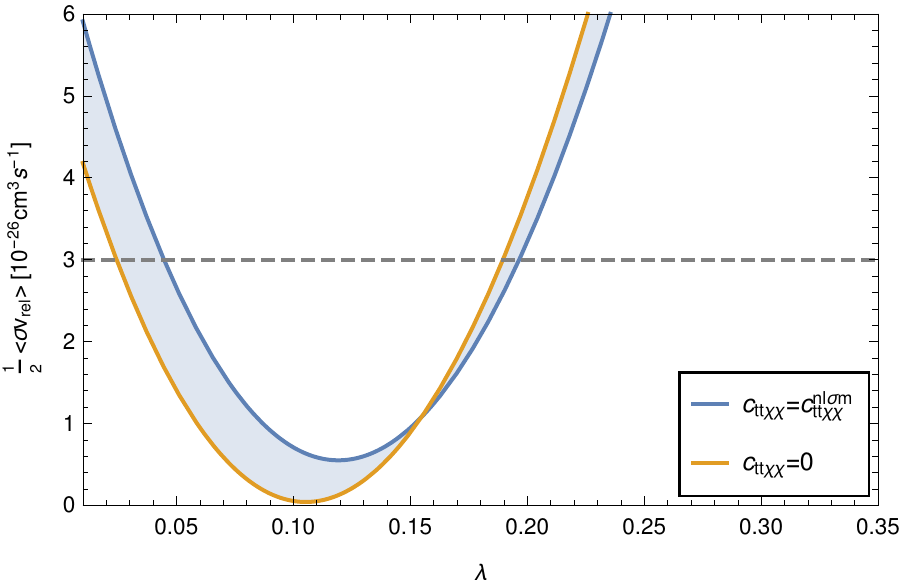}
\caption{Impact on the total annihilation cross section of varying the strength of the $t\bar{t}\chi^\ast \chi$ contact interaction in the range $c_{tt\chi\chi}\in [c_{tt\chi\chi}^{\rm nl\sigma m} , 0]$. The lower value corresponds to the pure sigma model, where top partner mixing is neglected, whereas the upper value corresponds to a setup with fully composite $t_R$, where top partner mixing is maximal. The realistic parameter points lie within this range, i.e. they fall within the band shaded in blue. The gray dashed line shows the value required to reproduce the present relic abundance according to the approximate relation in Eq.~\eqref{eq:DMrelic}. In the left panel we set $\lambda = 0.05$, whereas in the right panel the DM mass was fixed to $m_\chi = 300$ GeV. We took $f = 1.4$ TeV in both panels.}
\label{fig:thermXsectTPContr}
\end{figure}
For $m_\chi > m_t$, the simple scaling in Eq.~\eqref{eq:cancellation} is violated by the $\chi \chi^\ast \to t\bar{t}$ amplitude, where the $t\bar{t}\chi^\ast \chi$ contact interaction plays an important role. This is illustrated in Fig.~\ref{fig:thermXsectTPContr}, where we show the effect on the total annihilation cross section of varying the $t\bar{t}\chi^\ast \chi$ coupling within the range $c_{tt\chi\chi}^{\rm nl\sigma m} < c_{tt\chi\chi} < 0$, which contains all phenomenologically interesting points (recall Fig.~\ref{fig:TP_coupl_Corr}). The effect of top partner mixing is to suppress $|c_{tt\chi\chi}|$, which in turn shifts the DM relic abundance contour to larger $m_\chi$ for fixed $\lambda$, or conversely, to smaller $\lambda$ for fixed DM mass. As can be seen in the right panel of Fig.~\ref{fig:thermXsectTPContr}, at fixed $m_\chi$ the shift is larger for the branch with smaller $\lambda$. This can be explained by noticing that the size of the amplitude containing the $t\bar{t}\chi^\ast \chi$ contact interaction, relative to the one that couples $\chi \chi^\ast$ to an $s$-channel virtual Higgs via the portal coupling, is parametrically $2m_\chi^2 /(\lambda f^2)$ (for $m_\chi \gg m_h/2$). On the branch with larger $\lambda$ this ratio is smaller than $1$, so the corrections to the $t\bar{t}\chi^\ast\chi$ coupling play a subleading role. Conversely, on the branch with smaller $\lambda$ the ratio is larger than $1$, hence the reduction of the portal coupling caused by top partner mixing is sizable. As it will be shown below, this effect is crucial to evade direct detection bounds.

\subsection{Radiative corrections to pNGB derivative interactions} \label{sec:finitemom}
\begin{figure}[t]
\centering
\includegraphics[width=0.8\textwidth]{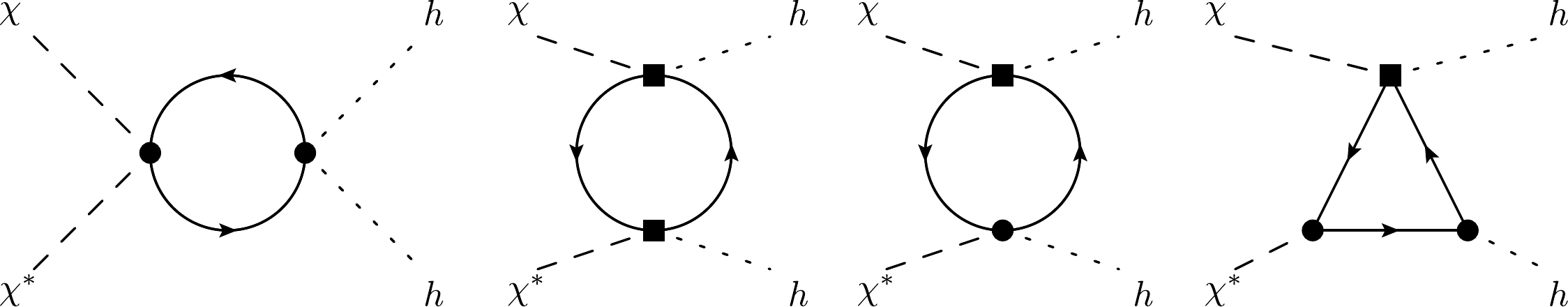}
\caption{Representative set of $1$-loop diagrams that contribute to the renormalization of the $\chi^\ast \chi hh$ interaction at $O(p^2)$. The circles indicate non-derivative interactions arising from elementary-composite mixing terms, whereas the squares denote derivative couplings originating from the $e_\mu$ symbol (see Eq.~\eqref{eq:fermion}).}
\label{fig:FiniteMomCorrections}
\end{figure}

Throughout our discussion thus far, the effects of gauge and fermionic loops were taken into account via the CW effective potential. In particular, for the computation of the annihilation cross sections we made use of Eq.~\eqref{eq:Leff_ann}, where the tree-level couplings were supplemented by the $1$-loop CW term. The effective potential, however, only captures the radiative corrections in the approximation of vanishing external momenta. This is not appropriate for DM annihilation, where the relevant external momentum scale is $p \sim m_\chi$, and $1$-loop corrections to derivative operators of $O(p^2)$ are expected to be also important. As an illustrative example, let us consider the $\chi\chi^\ast hh$ interaction at high energies, where EWSB effects can be neglected. From Eq.~\eqref{eq:LEGBint}, the tree-level (derivative) coupling reads simply
\begin{equation} \label{eq:derivtree}
\mathcal{L}_{\rm GB} \supset \frac{1}{f^2} h \partial_\mu h (\chi^\ast \partial^\mu \chi + \chi \partial^\mu \chi^\ast).
\end{equation}
Radiative corrections to this interaction arise only from the fermion sector. The $O(p^0)$ \mbox{$1$-loop} contribution is proportional to the $SO(7)$-breaking parameters $\epsilon$ and is just given by the portal coupling, $V_{\rm eff} \supset \lambda h^2 \chi^\ast \chi\,$. It is in general logarithmically UV-divergent, but it is rendered finite by the set of WSRs in Eq.~\eqref{eq:WSR1}. The $O(p^2)$ $1$-loop term must also be proportional to the $\epsilon$ parameters, because in the limit of vanishing explicit breaking, $\epsilon \to 0$, the $O(p^2)$ scalar Lagrangian is simply given by the sigma model kinetic term, Eq.~\eqref{eq:2deriv}, whose coefficient is fixed by $f$. Then the radiatively corrected form of the two-derivative coupling can be estimated as
\begin{equation}
i(c_{\rm tree} + c_{\rm 1-loop}) \frac{p^2}{f^2}\,,\qquad c_{\rm tree} \sim 1\,,\qquad c_{\rm 1-loop} \sim \frac{N_c \epsilon^2}{16\pi^2 f^2} \log \Lambda^2\,.
\end{equation}
Notice that $c_{\rm 1-loop}$ is expected to be logarithmically divergent, since the WSRs in Eqs.~(\ref{eq:WSR1},~\ref{eq:WSR2}) do not soften its UV behavior. The log enhancement, together with the fact that in general the ratio $\epsilon/f$ is of $O(1)$ or even somewhat larger, make this $1$-loop correction potentially very important and thus warrant a detailed calculation. We find four classes of diagrams that renormalize the operator in Eq.~\eqref{eq:derivtree}, depicted in Fig.~\ref{fig:FiniteMomCorrections}. Two types of fermion-scalar vertices appear in the diagrams: the non-derivative couplings arising from elementary-composite mixing terms, as well as the derivative couplings originating from the $e_\mu$ symbol that are contained in the kinetic terms of the resonances in the $SO(6)$ fundamental, $\sum_{i} \bar{Q}_i \slashed{e} Q_i\,$.\footnote{Notice that in general the couplings containing the $d_\mu$ symbol that appear in Eq.~\eqref{eq:Fermint} also contribute. However, for simplicity we set their coefficients to zero in the computation.} Neglecting external masses, so that $s + t + u \simeq 0$, we find for the $O(p^2)$ piece of the $\chi^\ast \chi \to hh$ amplitude (see App.~\ref{app:C} for details)
\begin{equation} \label{eq:finitemomResult}
i(c_{\rm tree} + c_{\rm 1-loop}) \frac{s}{f^2}\,,\qquad c_{\rm tree} = 1\,,\qquad c_{\rm 1-loop} = \frac{N_c}{2\pi^2 f^2}\Big(\epsilon_t^2 - \frac{\epsilon_q^2}{8}\Big) \log \frac{\Lambda^2}{m^2_\ast}\,,
\end{equation}
where we have imposed the WSRs, and $m_\ast$ stands for the mass of some fermionic resonance. Notice the mild loop suppression factor $N_c/(2\pi^2)$, and the log enhancement. After EWSB, this interaction contributes to the trilinear $\chi^\ast \chi h$ derivative coupling,
which as we discussed in Sec.~\ref{subsec:DMra} enters all annihilation cross section amplitudes, and in fact dominates in the viable region of parameters, where $\lambda$ is suppressed. Therefore in order to retain predictivity, we must keep the size of the radiative correction under control. We find an irreducible uncertainty of about $50\%$ at the cross section level, which corresponds to 
\begin{equation} \label{eq:FMcondition}
0.5 < \Big(1 + \frac{c_{\rm 1-loop}}{c_{\rm tree}}\Big)^2 < 1.5\qquad \longrightarrow \qquad -\, 0.4 < \frac{1}{f^2}\Big(\epsilon_t^2 - \frac{\epsilon_q^2}{8}\Big) < 0.3\,,
\end{equation}
where we have estimated $\Lambda \sim 10$ TeV and $m_\ast \sim 1$ TeV. Barring a cancellation $\epsilon_q^2 \approx 8\, \epsilon_t^2$, which may be regarded as a tuning unless it can be enforced by a symmetry, a further reduction of the uncertainty would lead to values of $\epsilon_{q,t}$ that are too small to reproduce the measured top mass. In conclusion, we will require that Eq.~\eqref{eq:FMcondition} is satisfied throughout our phenomenological analysis, and we will correspondingly assign a $50\%$ theoretical uncertainty on the total DM annihilation cross section.

\subsection{Constraints from DM direct detection}
Direct detection experiments aim at revealing DM-nucleus scattering events by measuring the nuclear recoil energy. Currently, the strongest constraints on the spin-independent (SI) DM-nucleon elastic cross-section come from the Xenon-based XENON1T \cite{Aprile:2017iyp} and LUX \cite{Akerib:2016vxi} experiments, with the former providing a slightly tighter bound. In our model, the elastic scattering of DM with a quark $q$ is mediated by three types of diagrams: Higgs exchange in the $t$-channel, the $\chi^\ast \chi \bar{q}q$ contact interaction, and diagrams involving the exchange of the $U(1)_{\rm DM}$-charged top partners $\mathcal{Y}, \mathcal{Z}$. The first two classes mediate scattering with all quarks, whereas the exchange of $\mathcal{Y},\mathcal{Z}$ only affects the scattering with (virtual) tops. Importantly, in Higgs exchange diagrams the contribution of the derivative coupling $\sim (v/f^2)\partial h \partial (\chi^\ast \chi)$ is suppressed by $-q^2/f^2 \ll 1$, where $\sqrt{-q^2} \lesssim 100$ MeV is the small momentum transfer. Therefore these diagrams are effectively proportional to the portal coupling $\lambda$. Furthermore, throughout the realistic parameter space the Higgs exchange amplitude dominates, being enhanced by $2 \lambda f^2 / m_h^2 \gg 1$ with respect to the sum of the other two terms. Hence the SI DM-nucleon cross section is well approximated by the simple expression familiar from the renormalizable Higgs portal model (see e.g. Ref.~\cite{Cline:2013gha}),
\begin{equation} \label{eq:DDxsection}
\sigma_{\rm SI}^{\chi N} \simeq \frac{f_N^2}{\pi} \frac{m_N^4 \lambda^2}{m_\chi^2 m_h^4} \;\sim\; 4 \cdot 10^{-46} \,\mathrm{cm}^2\, \left(\frac{\lambda}{0.03}\right)^2 \left(\frac{300\;\mathrm{GeV}}{m_\chi}\right)^2 ,
\end{equation}
where $m_N$ is the nucleon mass, and $f_N \simeq 0.30$ contains the dependence on the nucleon matrix elements. The exact expression of $\sigma_{\rm SI}^{\chi N}$ is reported in App.~\ref{app:C}. The cross section value $4\cdot 10^{-46}$ cm$^2$ corresponds to the current $90\%$ CL upper bound at $m_{\rm DM} = 300$ GeV from XENON1T \cite{Aprile:2017iyp}, showing that direct detection constraints require $\lambda$ to be suppressed by about a factor $2$ with respect to the most natural value $\lambda \sim \lambda_h/2\sim 0.065$. Notice that to calculate the excluded regions in the $(m_\chi, \lambda)$ plane of Fig.~\ref{fig:MainResults} below, the local DM density was assumed to take the standard value $\rho_0 = 0.3\;\mathrm{GeV}\, \mathrm{cm}^{-3}$, independently of the predicted thermal value. All direct detection constraints are given at $90\%$ CL.

\subsection{Results} \label{sec:results}
\begin{figure}[t]
  \includegraphics[scale=0.9]{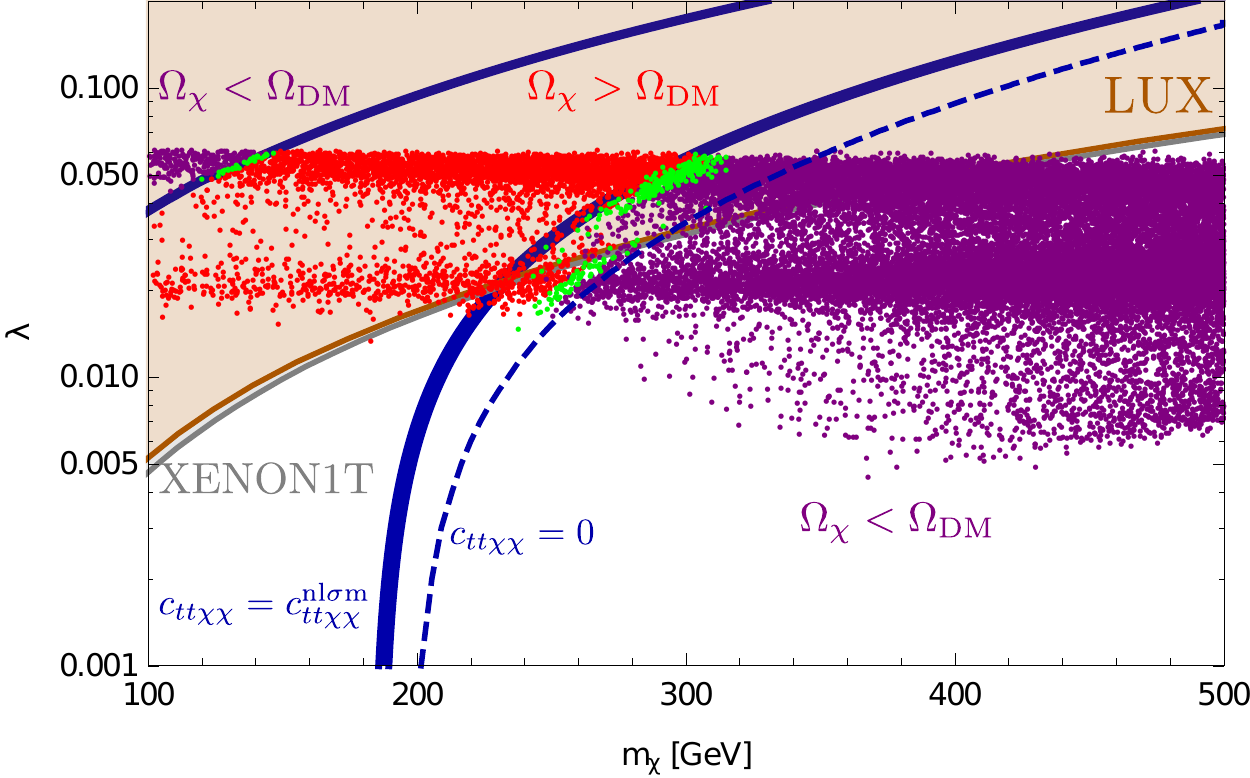}\par\vspace{1cm}
  \includegraphics[scale=0.9]{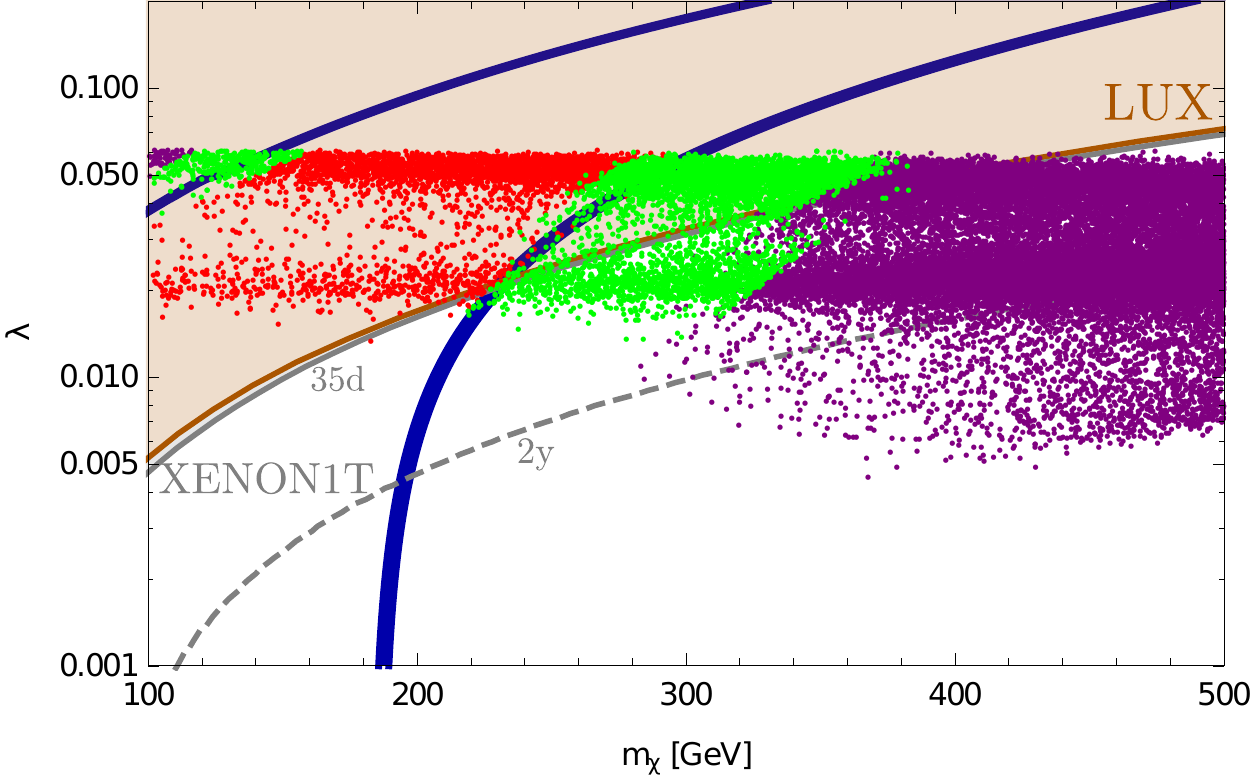}
\caption{Distributions in the $(m_\chi, \lambda)$ plane that summarize our analysis of DM phenomenology. The points have different colors depending on whether they are compatible with (green), exceed (red) or undershoot (purple) the observed value of the DM relic abundance. In the upper (lower) panel, the theoretical uncertainty of $50\%$ on the annihilation cross section is neglected (included). See the main text for further explanations on the meaning of the different curves.}
\label{fig:MainResults}
\end{figure}

The main results of our phenomenological analysis are shown in Fig.~\ref{fig:MainResults}. We set \mbox{$f = 1.4\;\mathrm{TeV}$} and perform a parameter scan, imposing that $v,\, m_h$ and $m_t$ match the experimental values. We also require each point to be compatible with detailed LHC constraints on top partners, which are discussed in Sec.~\ref{sec:TPbounds} below and summarized in Eq.~\eqref{eq:TPbounds}. In addition, the parameter space is restricted by the condition of Eq.~\eqref{eq:FMcondition}, thus ensuring that the theoretical uncertainty on the annihilation cross section due to missing radiative corrections is within $50\%$. The points are projected onto the plane $(m_{\chi},\lambda)$, using three different colors depending on whether the relic abundance is compatible with (green), exceeds (red) or undershoots (purple) the observed value. To compute the relic abundance we implemented the effective Lagrangian of Eq.~\eqref{eq:Leff_ann} in FeynRules \cite{Alloul:2013bka} and used micrOMEGAs \cite{Belanger:2013oya} to solve the Boltzmann equation (see App.~\ref{app:C} for details) for the DM density.\footnote{This treatment includes annihilation into light quarks and leptons, as well as into the three-body final states $WW^\ast$ or $ZZ^\ast$, which become important for lighter DM.} Notice that the couplings involving the top quark depend on the elementary-composite mixings and top partner masses, hence the relic abundance is not a function only of $f,\,m_\chi$ and $\lambda$, but must be separately evaluated at each point in parameter space. By contrast, the bounds from direct detection experiments, namely LUX (brown) and XENON1T (gray), are insensitive to the top partner parameters.

In the upper panel of Fig.~\ref{fig:MainResults}, we illustrate the effect of neglecting the $50\%$ theoretical uncertainty on $\langle \sigma v_{\rm rel} \rangle$, and show in green color the points that yield a relic abundance within $5\%$ of the observed value. For reference we also show, as thick blue lines, the $3\sigma$ relic abundance contours that are obtained by setting the $t\bar{t}h$ and $t\bar{t}\chi^\ast \chi$ couplings to their sigma model values. In this limit the annihilation cross section is completely fixed by $\{f, m_\chi, \lambda\}$. The ``two-branch'' structure discussed in Sec.~\ref{subsec:DMra} is clearly visible: for each value of \mbox{$m_\chi \gtrsim 180$} GeV there are two values of $\lambda$ that reproduce the correct relic abundance. In the upper branch DM annihilation proceeds dominantly through the portal coupling, whereas in the lower branch it is controlled primarily by the derivative interactions. In between the branches the two effects strongly cancel (see the discussion below Eq.~\eqref{eq:cancellation}), leading to a suppressed annihilation cross section and therefore to over-abundant DM. On the contrary, outside of the two branches one of the two couplings becomes too strong, and as a consequence the DM is under-abundant. The upper branch is robustly ruled out by direct detection, and we therefore focus on the lower branch. Here the green points fall between the two relic abundance contours obtained setting $c_{tt\chi\chi}= c_{tt\chi\chi}^{\rm nl\sigma m}$ (solid blue) and $c_{tt\chi\chi}= 0$ (dashed blue). The latter corresponds to maximal $t_R$ compositeness. For fixed $m_\chi$, a suppressed $|c_{tt\chi\chi}|$ reduces the portal coupling required for the correct relic abundance, and this in turn relaxes the direct detection constraints. Indeed, the subset of viable points that are compatible with direct detection limits lies close to the $c_{tt\chi\chi}= 0$ curve. Had we not included top partner mixing, we would have wrongly concluded that all these points are ruled out by LUX and XENON1T data. This highlights the importance of carefully taking into account the effects of the fermionic resonances.

In the lower panel of Fig.~\ref{fig:MainResults} we show the complete picture. The theoretical uncertainty is now included, so the green points reproduce the experimental value of the relic abundance within $50\%$. We find a large set of points that reproduce the relic abundance within the uncertainty, and at the same time evade the current direct detection bounds. The DM mass is in the range $200\;\mathrm{GeV} \lesssim m_\chi \lesssim 400\;\mathrm{GeV}$ and the portal coupling between roughly $0.01 \lesssim \lambda \lesssim 0.04$. We also show, as a dashed gray curve, the projected XENON1T sensitivity after two years of data taking \cite{Aprile:2015uzo} (whereas the ``35d'' label on the solid dashed curve refers to the current exposure of $35$ days \cite{Aprile:2017iyp}). All the currently viable points lie well within the ultimate reach of XENON1T, which will thus be able to test the entire parameter space of the model for $f= 1.4$ TeV.    

\subsection{Indirect detection}
\label{sec:indirect-det}
Indirect detection experiments, which search for signals of DM annihilation in the galaxy halo, constitute an additional probe of the model discussed here. Detailed constraints from the antiproton spectrum measured by PAMELA \cite{Adriani:2010rc} were presented, for the real singlet pNGB DM in the $SO(6)/SO(5)$ model, in Ref.~\cite{Marzocca:2014msa}. Since the annihilation pattern of our complex DM is very similar, we were able to check that the viable region of our parameter space is safely compatible with PAMELA antiproton data. It is important to observe that changing the assumptions on the systematic uncertainties that affect the astrophysical backgrounds can have a very large impact on the antiproton limits. For example, the more conservative approach taken in Ref.~\cite{Evoli:2015vaa} resulted in bounds on the DM annihilation cross section at present time, $\langle \sigma v_{\rm rel} \rangle_0$, that were an order of magnitude weaker than those quoted in Ref.~\cite{Marzocca:2014msa}. Very recently, Refs.~\cite{Cuoco:2016eej,Cui:2016ppb} used the new AMS-02 antiproton measurement \cite{Aguilar:2016kjl} to set very strong constraints. For example, assuming annihilation into $b\bar{b}$ the thermal value of the cross section $\langle \sigma v_{\rm rel} \rangle_0 \sim 3 \times 10^{-26}\;\mathrm{cm}^3 \,\mathrm{s}^{-1}$ was excluded for DM masses in the range $150\;\mathrm{GeV} \lesssim m_{\rm DM} \lesssim 500\; \mathrm{GeV}$ \cite{Cuoco:2016eej}. A detailed scrutiny of the AMS-02 constraints on pNGB DM, including the aforementioned large impact of the assumptions on systematic uncertainties, is an interesting direction for future work. Finally, we note that gamma ray observations of nearby dwarf spheroidal galaxies also set competitive bounds on DM annihilation, while being affected by smaller systematics compared to the antiproton channel. The current limits are roughly $\langle \sigma v_{\rm rel} \rangle_0 \lesssim 10^{-25}\;\mathrm{cm}^3 \,\mathrm{s}^{-1}$ for DM mass in the few hundred GeV range \cite{Fermi-LAT:2016uux}.\footnote{We thank A.~Urbano for illuminating discussions about indirect detection constraints.}

\section{Collider phenomenology} \label{sec:LHC}
In this section the collider phenomenology of the model is outlined, focusing on the signals of fermionic top partners at hadron colliders, which constitute the most sensitive probe. Nevertheless, before discussing this aspect in more detail we briefly touch upon other observables. Due to its pNGB nature, the Higgs boson couples to the other SM particles with strength that deviates at $O(v^2/f^2)$ from the SM predictions. In particular, the $hVV$ coupling ($V = W, Z$) is rescaled by a factor $c_V = \sqrt{1 - \xi}\,$. For our benchmark value $f = 1.4$ TeV, the deviation is of $\approx 1.5\%$, which is unaccessible at the LHC, but will be tested at future $e^+ e^-$ colliders (see Ref.~\cite{Durieux:2017rsg} for a recent overview). Parametrically similar deviations affect other SM couplings, such as $h\bar{t}t$, $hgg/h\gamma\gamma$ and $Z\bar{t}t$, which however will be tested with less accuracy than $hVV$. In addition, monojet searches only provide subleading constraints, because the coupling of $\chi$ to the proton constituents is very weak. In particular the contact interactions $\bar{q}q \chi^\ast \chi$, where $q$ is a light quark, are Yukawa-suppressed (if present at all). 

\subsection{LHC constraints on top partners} \label{sec:TPbounds}
A rather generic feature of pNGB Higgs models with partial compositeness is that the lightness of the Higgs requires at least some of the top partners to be light, $m_\ast = g_\ast f$ with $g_\ast \sim 1$ (see Ref.~\cite{Panico:2012uw} for an extensive discussion). In our model, this is illustrated by Fig.~\ref{fig:2layerhmass}. Since the top partners are colored, the searches for their signals at hadron colliders, in particular at the LHC, are among the most important experimental tests of the composite Higgs framework \cite{Contino:2006nn,Contino:2008hi,Mrazek:2009yu,DeSimone:2012fs}. In the following discussion we adopt a simplified model where only one layer of resonances, containing one $SO(6)$ fundamental $Q$ and one singlet $S$, is included. This captures the main phenomenological features of the complete model, provided the second layer of resonances is somewhat heavier than the first, as it is the case in most of the parameter space.

\begin{figure}[t]
\includegraphics[width=0.495\textwidth]{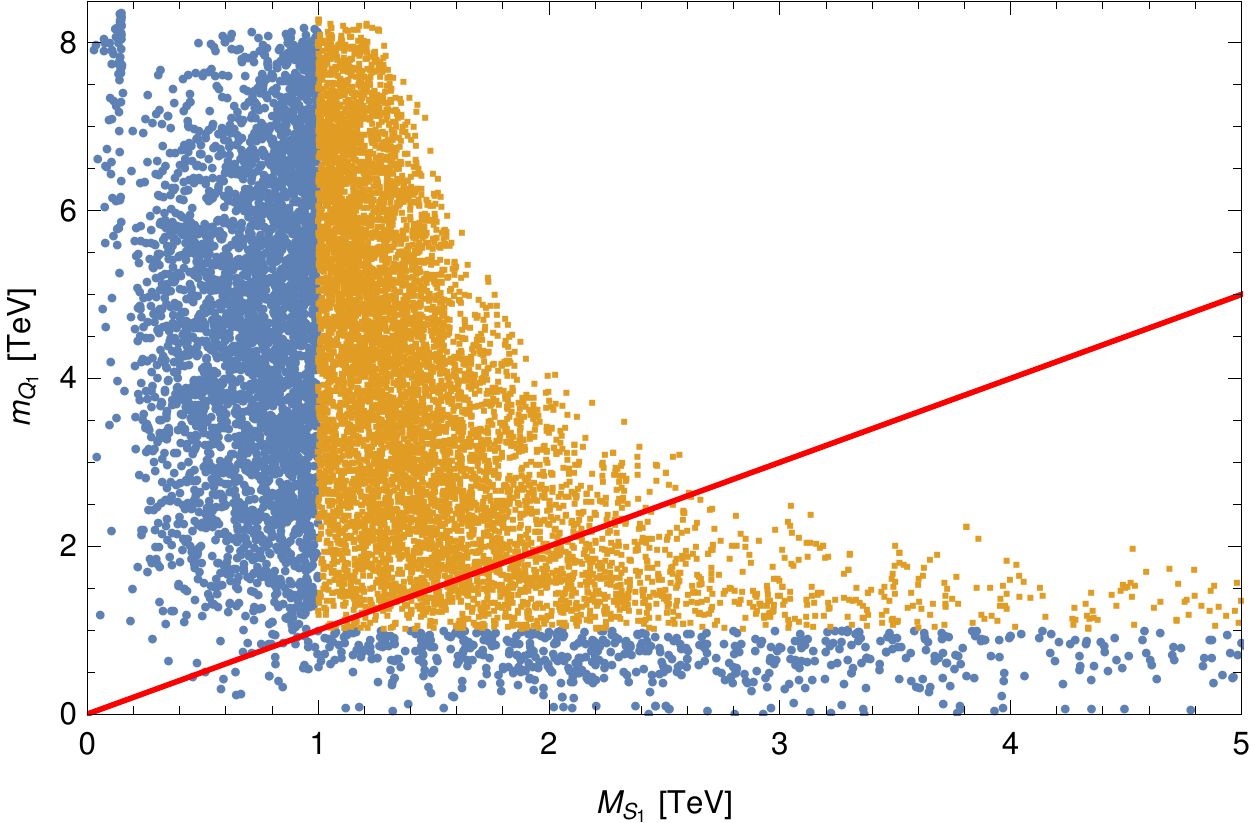}
\includegraphics[width=0.495\textwidth]{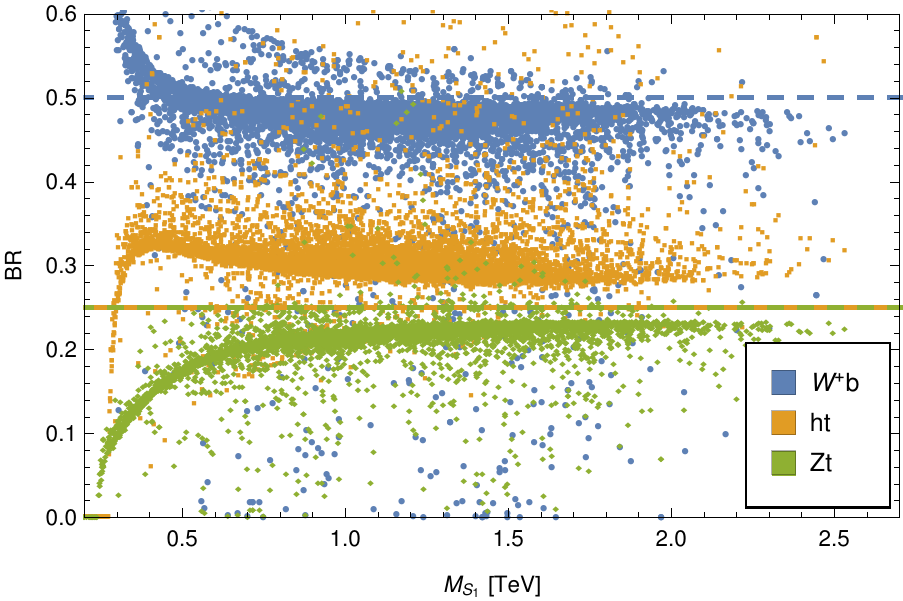}
\caption{Distributions in the model with two fermionic resonance layers. {\it Left:} mass of the lightest exotic top partner $(m_{Q_1})$ versus the mass of the lightest singlet top partner $(M_{S_1})$. For orange (blue) points, the lightest fermionic resonance is heavier (lighter) than the approximate LHC lower bound of $1$ TeV. The red line corresponds to $m_{Q_1} = M_{S_1}$. {\it Right:} branching ratios of the lightest singlet $S_1$, for the parameter points where it is the lightest fermionic resonance. The dashed lines indicate the leading order predictions, see Eq.~\eqref{eq:TPbr}.}
\label{fig:TPscans}
\end{figure}

We start from the fermionic Lagrangian in Eq.~\eqref{eq:fermion} with $N_Q = N_S = 1$. Notice that, as consistently done throughout our analysis, the coefficients of the derivative interactions in Eq.~\eqref{eq:Fermint} are set to zero, $c^{L,R} = 0$. We will return to the possible role of these interactions in LHC physics in Sec.~\ref{Sec:TPsignals}. Neglecting EWSB effects, the elementary-composite mixings are diagonalized by the rotations
\begin{equation} \label{eq:TProtations}
\begin{pmatrix} t_R \\ S_R \end{pmatrix} \to \begin{pmatrix} \cos \phi_R & - \sin \phi_R \\ \sin\phi_R & \cos\phi_R \end{pmatrix} \begin{pmatrix} t_R \\ S_R \end{pmatrix},\qquad \begin{pmatrix} q_L \\ Q_L \end{pmatrix} \to \begin{pmatrix} \cos \phi_L & - \sin \phi_L \\ \sin\phi_L & \cos\phi_L \end{pmatrix} \begin{pmatrix} q_L \\ Q_L \end{pmatrix},
\end{equation}
where $Q\equiv (T, B)^T$ and the mixing angles are $\tan\phi_R = \epsilon_{tS}/m_S$ and $\tan\phi_L = \epsilon_{qQ}/m_Q\,$. On the other hand, the remaining fermions contained in $Q$, namely the exotic doublet $(X_{5/3}, X_{2/3})^T$ and the $U(1)_{\rm DM}$-charged SM singlets $\mathcal{Y},\mathcal{Z}$, do not mix with the elementary fermions. In summary, the top partner masses are
\begin{equation}
M_S = \sqrt{m_S^2 + \epsilon_{tS}^2}\,,\qquad M_{T,\,B} = \sqrt{m_Q^2 + \epsilon_{qQ}^2}\,,\qquad M_{X_{5/3},\, X_{2/3},\, \mathcal{Y},\, \mathcal{Z}} = m_Q\,.
\end{equation}
Hence at the bottom of the spectrum we find either a singlet $S$, or four approximately degenerate states $X_{2/3}, X_{5/3}, \mathcal{Y}$ and $\mathcal{Z}$.\footnote{EWSB effects do not alter the masses of $X_{5/3}, \mathcal{Y}$ and $\mathcal{Z}$, which remain exactly degenerate, but they do shift $M_{X_{2/3}}$ slightly. The correction can have either sign depending on the parameter point.} The scan of the complete two-layer model, shown in the left panel of Fig.~\ref{fig:TPscans}, demonstrates that the lightest top partner is typically a singlet, although the alternative configuration is also possible. The decay patterns of the resonances can be immediately understood using the Goldstone equivalence theorem. Expanding the $U$ matrix to $O(1/f)$ and diagonalizing the elementary-composite mixings via Eq.~\eqref{eq:TProtations}, one immediately finds the leading order results
\begin{equation}
\begin{split}
\text{BR}(S\rightarrow W^+ b)&\, = 2\, \text{BR}(S\rightarrow Z t) = 2\, \text{BR}(S\rightarrow h t) = \frac{1}{2}\,,\\
\text{BR}(T\rightarrow h t)&\, = \text{BR}(T\rightarrow Z t) = \text{BR}(X_{2/3}\rightarrow h t) = \text{BR}(X_{2/3}\rightarrow Z t) = \frac{1}{2}\,,\\ 
\text{BR}(X_{5/3}\rightarrow W^+ t)&\, = \text{BR}(B\rightarrow W^- t) = \text{BR}(\mathcal{Y}\rightarrow \chi t) = \text{BR}(\mathcal{Z} \rightarrow \chi^* t) = 1\,.
\end{split}
\label{eq:TPbr}
\end{equation}
In particular, as a consequence of $U(1)_{\rm DM}$ conservation, $\mathcal{Y}\, (\mathcal{Z})$ always decays into a top quark and a $\chi\,(\chi^\ast)$ particle (see Refs.~\cite{Serra:2015xfa,Anandakrishnan:2015yfa,Chala:2017xgc} for recent studies of top partner decays into additional Goldstone scalars). The above predictions are well respected in the complete model. For example, in the right panel of Fig.~\ref{fig:TPscans} the exact branching ratios of the singlet are shown, for the parameter points where it is the lightest fermionic resonance. We find good agreement with Eq.~\eqref{eq:TPbr}.

\enlargethispage{-30pt}
The LHC searches for top partners target two distinct production mechanisms: pair production via the QCD interactions, namely $pp\to \bar{\psi}\psi$ where $\psi$ is a generic top partner, and single production in association with a top or bottom via the electroweak interactions, for example for a singlet $S$ the leading process is $pp \to S\bar{b}j$ via the $\bar{b}W^- S$ vertex. Notice that the $U(1)_{\rm DM}$-charged top partners $\mathcal{Y}$ and $\mathcal{Z}$ cannot be singly produced. We have verified that under our assumption $c_{ji}^{L,R} = 0$, the bounds from single production \cite{CMS:2017oef} are weaker than those coming from QCD pair production \cite{ATLAS:2016btu,CMS:2017jfv}, hence we only discuss the latter. For simplicity, in the following we set the branching ratios to the approximate values of Eq.~\eqref{eq:TPbr}. The search of Ref.~\cite{ATLAS:2016btu} focuses on the $\bar{\psi}\psi \to t(h \to b\bar{b})$+$X$ process in $1$- and $0$-lepton final states, yielding the $95\%$ CL constraints $M_S > 1.02$ TeV and $M_{X_{2/3}} > 1.16$ TeV (henceforth, LHC limits will always be quoted at $95\%$ CL). The bound on $X_{2/3}$ is stronger due to the larger branching ratio into $th$. The search of Ref.~\cite{CMS:2017jfv} instead specifically targets the $X_{5/3}$ in the same-sign-dileptons final state, and gives $M_{X_{5/3}} > 1.16$ TeV.\footnote{This is the bound obtained for a purely right-handed $\bar{t}\,W^- X_{5/3}$ coupling, as appropriate since in this model the left-handed coupling is suppressed by one extra power of $v$.} In addition to these ``standard'' constraints, we must account for those on $\mathcal{Y}$ and $\mathcal{Z}$, which are mass-degenerate and always decay into a top quark and a DM particle, giving rise to $t\bar{t}$ + missing transverse energy (MET) signatures. The corresponding constraint depends on the DM mass. As motivated by the results of our phenomenological analysis (see Fig.~\ref{fig:MainResults}), we choose the representative value $m_\chi = 300$ GeV. To estimate the current bound on $M_{\mathcal{Y}} = M_{\mathcal{Z}}$, we start from the result obtained in the dedicated $8$ TeV analysis of Ref.~\cite{Kraml:2016eti}, $m_\psi > 0.85$ TeV based on $\sim 20$ fb$^{-1}$ of data. Using the Collider Reach \cite{ColliderReach} method, we rescale this bound to the current luminosity and energy, $\sim 36$ fb$^{-1}$ at $13$ TeV, obtaining $m_\psi > 1.30$ TeV. Finally, to take into account that $\mathcal{Y}$ and $\mathcal{Z}$ are {\it two} degenerate Dirac fermions that contribute to the signal, we solve the following equation for $M_\mathcal{Y}\,$: $\sigma_{pp\,\to\, \bar{\psi}\psi,\,13\; \mathrm{TeV}}(m_\psi = 1.30\;\mathrm{TeV}) = 2\,\sigma_{pp\,\to\, \bar{\psi}\psi,\,13\; \mathrm{TeV}}(M_\mathcal{Y})$, arriving to $M_\mathcal{Y} > 1.42$ TeV.\footnote{As an independent cross-check, we have recast the constraint on the stop mass extracted from Ref.~\cite{Sirunyan:2017kqq}, $m_{\tilde{t}} > 1.04$ TeV with $\sim 36$ fb$^{-1}$, by solving for $M_{\mathcal{Y}}$ the equation \mbox{$\sigma_{pp\,\to\, \tilde{t}^\ast \tilde{t},\,13\; \mathrm{TeV}}(m_{\tilde{t}} = 1.04\;\mathrm{TeV}) = 2\,\sigma_{pp\,\to\, \bar{\psi}\psi,\,13\; \mathrm{TeV}}(M_\mathcal{Y})$}, obtaining a consistent bound $M_\mathcal{Y} > 1.47$ TeV.} In summary, the current LHC constraints on the top partner masses are, at $95\%$ CL,
\begin{equation} \label{eq:TPbounds}
M_S > 1\;\mathrm{TeV},\;\qquad M_{X_{5/3},\,X_{2/3}} > 1.2\;\mathrm{TeV},\;\qquad M_{\mathcal{Y},\,\mathcal{Z}} > 1.4\;\mathrm{TeV}.
\end{equation}
These conditions are imposed at every point in the parameter scan presented in Fig.~\ref{fig:MainResults}.

\subsection{Beyond the lightest top partner(s)}\label{Sec:TPsignals}
As discussed in Sec.~\ref{sec:TPbounds}, the first experimental manifestation of the model at colliders would most likely be the discovery of the lightest top partner. We now turn to a brief discussion of the opportunities to probe the heavier fermionic resonances at the LHC and future colliders. If the lightest top partner is a singlet $S$, the connection with the DM problem could not be made until the $U(1)_{\rm DM}$-charged top partners $\mathcal{Y}$ and $\mathcal{Z}$, which belong to the heavier multiplet $Q$, can be accessed. For large enough splitting $m_Q - M_S$, the direct decay to $\chi^{(\ast)} t$ and the cascade decay to $\chi^{(\ast)} S$ are both unsuppressed. The branching fraction is, assuming $\epsilon_{tQ}, M_S \ll m_Q$ and in the limit of full $t_R$ compositeness $\sin\phi_R \to 1$,
\begin{equation} \label{eq:YBR}
\mathrm{BR}(\mathcal{Y} \to \chi t) = \mathrm{BR}(\mathcal{Z} \to \chi^\ast t ) \simeq \frac{c_R^2}{c_L^2 + c_R^2}\,,
\end{equation}
where to keep the discussion general we took nonzero coefficients for the derivative interactions in Eq.~\eqref{eq:Fermint}, setting $c^{L,R} = i \,c_{L,R}$ so that $CP$ is conserved. Equation \eqref{eq:YBR} suggests that $\mathcal{Y}$ and $\mathcal{Z}$ decay rather democratically into the two available channels. Therefore the QCD pair production of $\mathcal{Y}$ and $\mathcal{Z}$, either at the LHC or at a future FCC-hh, can generate cascades where the decay of an intermediate $S$ yields a $Z$ or $h$ in addition to the ``stop-like'' $bW\bar{b}W \chi \chi^\ast$ signature, potentially providing an additional handle to characterize the exotic top partners.

In the opposite scenario $m_Q < M_S$, since the $\mathcal{Y}$ and $\mathcal{Z}$ are at the bottom of the spectrum, their discovery in the $t\bar{t}$ + MET final state would happen early on, hinting to a connection with DM physics. The heavier singlet may then be accessed via single production $pp\to S \bar{b} j$, whose rate can be enhanced by the derivative interactions proportional to $c_{L,R}$ \cite{DeSimone:2012fs}. Of special interest is the decay into the $U(1)_{\rm DM}$-charged top partners, $S\to \chi^\ast \mathcal{Y}, \chi \mathcal{Z}$, leading at the end of the cascade to the final state $t \chi \chi^\ast \bar{b}j$, i.e. a monotop signature. The branching ratio for these decays is, assuming $\epsilon_{tQ}, m_Q \ll M_S$ and in the limit of full $t_R$ compositeness,
\begin{equation}
\mathrm{BR}(S \to \chi^\ast \mathcal{Y}) = \mathrm{BR}(S \to \chi \mathcal{Z}) \simeq \frac{1}{6}\,.
\end{equation}
Notice that this result holds for arbitrary $c_L$ and $c_R$. Hence $\approx 1/3$ of the singly-produced singlets yield the monotop final state. This promising signature deserves a dedicated analysis, which is however beyond the scope of this paper.

\section{Outlook} \label{sec:outlook}
We have presented a model where a composite pNGB scalar DM $\chi$ is charged under an exact $U(1)_{\rm DM}$ that belongs to the unbroken symmetry group of the strong sector. This provides a robust stabilization mechanism for $\chi$, since the $U(1)_{\rm DM}$ is automatically respected by any $\mathcal{G}$-invariant UV completion. Higher-dimensional derivative operators that arise from the nonlinear sigma model play an important role in the DM phenomenology  \cite{Frigerio:2012uc}. They can give a large contribution to DM annihilation in the early Universe, while at the same time yielding negligible effects in the scattering with the nuclei of underground detectors. As a consequence, the tension with the strong direct detection constraints on the marginal coupling $\sim \lambda \chi^\ast \chi H^\dagger H$ is relaxed compared to the standard Higgs portal model, where $\lambda$ is fixed by the relic density. An extensive phenomenological analysis reveals a region of parameters compatible with all current constraints from the relic density, direct detection experiments and LHC searches for the top partners. The symmetry breaking scale is $f = 1.4$ TeV, the mass of the DM $200\;\mathrm{GeV}\lesssim m_\chi \lesssim 400$ GeV and the portal coupling \mbox{$0.01 \lesssim \lambda \lesssim 0.04$}. We found that the mixing of the top quark with the top partners plays a crucial role in obtaining the correct relic abundance, while at the same time evading the current direct detection constraints. In addition, we have identified a set of radiative corrections to the derivative operators, that imply a sizable theoretical uncertainty on the DM annihilation cross section and therefore a broadening of the allowed region of parameters. Nevertheless, this DM candidate falls within the ultimate sensitivity of XENON1T, and therefore will be fully tested in the near future. Indirect detection constraints and the impact of the assumed systematic errors are an interesting direction for future work.

Throughout the discussion, we have assumed that the $U(1)_{\rm DM}$ is a global symmetry. We wish to conclude with a few remarks about the possibility of weakly gauging it, with associated gauge boson $A_{D}^\mu$ and coupling $g_{D}$. To preserve the DM stability, we assume that $U(1)_{\rm DM}$ is exact, and therefore $A_D$ is massless. An immediate concern is the kinetic mixing with the SM hypercharge gauge boson, $(\epsilon/2) B_{\mu\nu} F_D^{\mu\nu}$, which after diagonalization of the kinetic terms leads to a small hypercharge for the DM $\chi$, hence in particular a coupling of size $\sim \epsilon g_D$ to the SM photon. Millicharged DM can remain tightly coupled to the baryon-photon plasma during the recombination epoch, behaving like a baryon and thus affecting the CMB. Requiring that the DM decouples from the plasma before recombination yields a constraint $\epsilon (g_{D}/e) \lesssim 5 \times 10^{-5}$ for $m_\chi = O(100)$ GeV \cite{McDermott:2010pa}. In addition, strong bounds from direct detection arise, but these do not apply in a wide region of $\epsilon g_D$ where the DM would have been evacuated from the galactic disk by supernova explosion shock waves, and prevented to return by galactic magnetic fields \cite{Chuzhoy:2008zy,McDermott:2010pa}. In our model, composite resonances charged under both $U(1)_Y$ and $U(1)_{\rm DM}$ do exist, for example the fermionic top partners $\mathcal{Y}$ and $\mathcal{Z}$, as well as gauge resonances. However, at $1$-loop no kinetic mixing is induced, because for each multiplet $\Phi \sim \mathbf{R}_X$ under $SO(6)\times U(1)_X$ we have $\epsilon^{\rm 1-loop}_{\Phi} \propto \mathrm{Tr}\,[(T_R^3 + X) T^{\rm DM}] = 0$, where $T_R^3$ and $T^{\rm DM}$ are $SO(6)$ generators in the $\mathbf{R}$ representation. In addition, a preliminary investigation suggests that $\epsilon$ may be even further suppressed, due to the non-abelian nature of $T^{\rm DM}\in SO(6)$. If the kinetic mixing is small enough -- or altogether absent, -- then DM charged under a dark $U(1)$ has been shown to be broadly compatible with astrophysical observations \cite{Feng:2008mu,Ackerman:mha,Feng:2009mn,Agrawal:2016quu}. In particular, the recent reappraisal of Ref.~\cite{Agrawal:2016quu} suggested the viability of charged DM with weak scale mass and coupling to the dark photon of strength only moderately weaker than electromagnetism. In the light of the above considerations, the effects of gauging the $U(1)_{\rm DM}$ on the DM phenomenology discussed in this paper are expected to be mild, because for weak-scale $m_\chi$ the dominant interactions are set by larger couplings, especially the top Yukawa. On the other hand, in the scenario where the mixings of the top with the strong sector preserve the $\chi$ shift symmetry, leading to $m_\chi \ll m_h$ and suppressed DM-SM couplings, the additional interactions with the dark photon can lead to significant modifications of the phenomenology \cite{LightDMfuture}.

\vspace{3.5cm}
\noindent {\bf Acknowledgments} 

We are grateful to R.~Harnik, A.~Ibarra, A.~Katz, A.~Pomarol, F.~Sala and A.~Urbano for useful conversations, and to J.~Serra for valuable comments about the manuscript. This work has been partially supported by the DFG Cluster of Excellence 153 ``Origin and Structure of the Universe,'' by the Collaborative Research Center SFB1258, the COST Action CA15108, and the European Union''s Horizon 2020 research and innovation program under the Marie Curie grant agreement, contract No.~675440. The work of RB is supported by the Minerva foundation. RB, ES and AW thank the Mainz Institute for Theoretical Physics for hospitality and partial support during the completion of this work.
\appendix
\section{CCWZ construction for $SO(7)/SO(6)$} \label{app:A}
For the generators of the fundamental representation of $SO(7)$ we take
\begin{align}
(T^{\alpha}_{L, R})_{IJ} &\,=\, -\frac{i}{2}\left[\frac{1}{2}\epsilon^{\alpha\beta\gamma}	(\delta^\beta_I \delta^\gamma_J - \delta^\beta_J \delta^\gamma_I ) \pm (\delta^\alpha_I \delta^4_J - \delta^\alpha_J \delta^4_I )\right], \qquad \alpha = 1, 2, 3, \nonumber \\
T_{IJ}^{ab} &\,=\, -\frac{i}{\sqrt{2}} (\delta^a_I \delta^b_J - \delta^a_J \delta^b_I),\qquad b=5,6;\; a = 1, \ldots, b -1,\\
X_{IJ}^{a} &\,=\, -\frac{i}{\sqrt{2}}(\delta^{a}_{I} \delta^7_{J} -	\delta^{a}_{J}	\delta^7_{I}),\qquad a = 1,\ldots,6, \nonumber
\end{align}
where the indices $I,J$ take the values $1, \ldots, 7$. $T^{\alpha}_{L, R}$ and $T^{ab}$ are the generators of $SO(6)$, collectively denoted by $T^{\hat{a}}$ ($\hat{a} = 1, \ldots, 15$), with $T^{\alpha}_{L, R}$ spanning the custodial \mbox{$SO(4)\cong SU(2)_L\times SU(2)_R$} subgroup, while $X^{a}$ are the broken generators that parameterize the coset space $SO(7)/SO(6)$. Notice that the unbroken generators are block-diagonal in our basis, 
\begin{equation}
T^{\hat{a}} = \begin{pmatrix} t^{\hat{a}} & 0 \\ 0 & 0 \end{pmatrix}, \qquad t^{\hat{a}} \in SO(6).
\end{equation}
All generators $T^A$ ($A=1,\ldots,21$) are normalized such that $\text{Tr}\left[T^A T^B\right]=\delta^{AB}$. Under the unbroken $SO(6)$, the six GBs $\pi^{a}$ transform linearly and in the fundamental representation, whose decomposition under $SO(4)$ is $\mathbf{6} = \mathbf{4} \oplus \mathbf{1} \oplus \mathbf{1}$. The Higgs doublet $H = \left( h_u , h_d\right)^T$ is identified with the $\mathbf{4}$, so that
\begin{equation}
\vec{\pi} = \frac{1}{\sqrt{2}}\begin{pmatrix}
-i(h_u -h_u^\ast ), &
h_u + h_u^\ast, & 
i( h_d - h_d^\ast ), &
h_d + h_d^\ast, &
\sqrt{2} \,\eta, & 
\sqrt{2} \,\kappa
\end{pmatrix}^T.
\label{eq:GBEmbedding}
\end{equation}
In unitary gauge, i.e. $h_u=0$, $h_d= \tilde{h} / \sqrt{2}$, this has the expression in Eq.~\eqref{eq:pionsUG} and the Goldstone matrix $U (\vec{\pi}) = \exp \left( i\sqrt{2} \pi^{a} X^{a}/f \right)$ can be written, after performing the convenient field redefinition \cite{Gripaios:2009pe}
\begin{equation} \label{eq:fieldRedef}
\frac{\sin (\pi/f)}{\pi}\,\pi^{a}\rightarrow \frac{\pi^{a}}{f}\qquad \text{with }\quad \pi=\sqrt{\vec{\pi}^{\,T} \vec{\pi}}\,,
\end{equation}
in the following form
\begin{equation}
U = \begin{pmatrix}
\mathbf{1}_{3\times 3} & & & &\\
& 1-\frac{\tilde{h}^2}{f^2 (1+\Omega)} & -\frac{\tilde{h}\eta}{f^2 (1+\Omega)} & -\frac{\tilde{h} \kappa}{f^2 (1+\Omega)}&\frac{\tilde{h}}{f}\\
& -\frac{\tilde{h} \eta}{f^2 (1+\Omega)} & 1-\frac{\eta^2}{f^2 (1+\Omega)} & -\frac{\eta \kappa}{f^2 (1+\Omega)}&\frac{\eta}{f}\\
& -\frac{\tilde{h}\kappa}{f^2 (1+\Omega)} & -\frac{\eta \kappa}{f^2 (1+\Omega)} & 1-\frac{\kappa^2}{f^2 (1+\Omega)}&\frac{\kappa}{f}\\
& -\frac{\tilde{h}}{f} & -\frac{\eta}{f} & -\frac{\kappa}{f}& \Omega
\end{pmatrix}\,,\qquad \Omega = \frac{1}{f}\sqrt{f^2 - \tilde{h}^2 - \eta^2 - \kappa^2}\,.
\label{eq:GBMatrix}
\end{equation}
Under $g\in SO(7)$, the GB matrix transforms as
\begin{equation}
U\left(\vec{\pi}\right) \to g \, U\left(\vec{\pi}\right) \, h\left(\vec{\pi};g\right)^T ,
\end{equation}
where $h\left(\vec{\pi};g\right)$ is block-diagonal in our basis,
\begin{equation}
h\left(\vec{\pi};g\right) = \begin{pmatrix}
h_6	&	0\\
0	&	1
\end{pmatrix}, \qquad h_6\in SO(6).
\label{eq:SO6embedding}
\end{equation}
The $d_\mu$ and $e_\mu$ symbols are defined via\footnote{Notice that we define the $d_\mu$ and $e_\mu$ symbols with opposite sign compared to, e.g., Ref.~\cite{DeSimone:2012fs}.}
\begin{equation}
iU^\dagger D_\mu U \equiv d_\mu^{a} X^{a} + e_\mu^{\hat{a}} T^{\hat{a}}\,,
\end{equation}
where $D_\mu U = \partial_\mu U - i A_\mu^{\hat{a}} T^{\hat{a}} U$. Notice that we took the gauge fields as belonging to the $SO(6)$ subalgebra, since this is the relevant case. Explicitly, for $SU(2)_L \times U(1)_Y$ we have $A_\mu^{\hat{a}} T^{\hat{a}} = \bar{g} \bar{W}_\mu^{\alpha} T_L^{\alpha}  + \bar{g}' \bar{B}_\mu T^{3}_R\,$. If the $U(1)_{\rm DM}$ were also gauged, then $A_\mu^{\hat{a}} T^{\hat{a}} \to A_\mu^{\hat{a}} T^{\hat{a}} +  g_{D} A_{D\mu} \sqrt{2}\, T^{56}$, with $A_{D}$ the associated vector field and $g_D$ its coupling. Under $g\in SO(7)$, 
\begin{equation}
d_\mu^{a} \to  \left(h_6\right)^{a}_{\;\,b} d_\mu^{b}\,, \qquad 
e_\mu \equiv e_\mu^{\hat{a}} t^{\hat{a}} \to  h_6 \left(e_\mu + i\partial_\mu\right) h_6^T\,,
\end{equation}
where $h_6$ was defined in Eq.~(\ref{eq:SO6embedding}). To leading order in $1/f$, we have
\begin{equation}
d_\mu^{a} = - \frac{\sqrt{2}}{f}\, D_\mu \pi^{a} + O(1/f^3), \qquad e_\mu^{\hat{a}} =  A_{\mu}^{\hat{a}} + O(1/f^2)\,,
\end{equation}
where $D_\mu \pi^{a} = \partial_\mu \pi^a - i A_\mu^{\hat{a}} (t^{\hat{a}})^{a}_{\;\,b} \pi^b\,$. The fermion covariant derivatives that appear in Eq.~\eqref{eq:fermion} read
\begin{equation}
D_\mu q_L = \Big(\partial_\mu - i \bar{g} \bar{W}^{\alpha}_{\mu} \frac{\sigma^\alpha}{2} - i \bar{g}' \frac{1}{6} \bar{B}_\mu\Big) q_L\,, \qquad D_\mu \Psi = \Big( \partial_\mu - i \frac{2}{3} \bar{g}' \bar{B}_\mu \Big) \Psi\,,
\end{equation}
where $\Psi = t_R, Q_i, S_j\,$, and in all cases the color $SU(3)$ component is understood.

At the leading order in derivatives, the Lagrangian describing the vector resonances $\rho_\mu \equiv \rho_\mu^{\hat{a}} t^{\hat{a}} \sim \mathbf{15}$ and $a_\mu \equiv a_\mu^{a} X^{a} \sim \mathbf{6}$ reads
\begin{equation}
\mathcal{L}_{V} =  -\frac{1}{4}\text{Tr}\left( \rho_{\mu\nu} \rho^{\mu\nu}\right) + \frac{f_{\rho}^2}{2}\text{Tr}\left(g_{\rho} \rho_\mu - e_\mu \right)^2 -\frac{1}{4}\text{Tr}\left( a_{\mu\nu}a^{\mu\nu}\right) + \frac{f_{a}^2}{2 \Delta^2}\text{Tr}\left(g_{a} a_\mu - \Delta d_\mu\right)^2\,,
\label{eq:vectorRes}
\end{equation} 
where $f_{\rho,\, a}$ are decay constants, $g_{\rho, \,a}$ are couplings, and $\Delta$ is a dimensionless parameter. The field strengths are given by
\begin{equation}
\rho_{\mu\nu} = \partial_\mu \rho_\nu - \partial_\nu \rho_\mu - i g_{\rho} \left[ \rho_\mu,\rho_\nu\right],\qquad a^{a}_{\mu\nu}=\nabla_\mu a^{a}_\nu - \nabla_\nu a^{a}_\mu, \quad \nabla_\mu = \partial_\mu - i e_\mu\,.
\end{equation}
In the limit where the external gauge fields are neglected, the masses of the $\rho$ and $a$ read
\begin{equation}
m_{\rho}^2 =  g_{\rho}^2 f_{\rho}^2 \,,\qquad m_{a}^2 = \frac{g_{a}^2 f_{a}^2}{\Delta^2}  \,.
\end{equation}
Neglecting EWSB, only $\rho_\mu$ can mix with the $SU(2)_L\times U(1)_Y$ gauge fields. The mass eigenstates are obtained via the rotations
\begin{equation}
\begin{pmatrix} \bar{W}^{\alpha} \\ \rho_{L}^\alpha \end{pmatrix} \to \frac{1}{\sqrt{g_\rho^2 + \bar{g}^2}} \begin{pmatrix} g_\rho & - \bar{g} \\ \bar{g} & g_\rho \end{pmatrix}  \begin{pmatrix} W^\alpha \\ \rho_L^{\alpha} \end{pmatrix},\qquad 
\begin{pmatrix} \bar{B} \\ \rho_{R}^3 \end{pmatrix} \to \frac{1}{\sqrt{g_\rho^2 + \bar{g}^{\prime\,2}}} \begin{pmatrix} g_\rho & - \bar{g}' \\ \bar{g}' & g_\rho \end{pmatrix}  \begin{pmatrix} B \\ \rho_R^{3} \end{pmatrix},
\label{eq:vectorMix}
\end{equation}
with $W^\alpha$ and $B$ identified with the SM states. The associated SM couplings are $g = g_\rho \bar{g} / \sqrt{g_\rho^2 + \bar{g}^2}$ and $g' = g_\rho  \bar{g}'  / \sqrt{g_\rho^2 + \bar{g}^{\prime\,2}}$.

\section{Scalar potential and parameter scan} \label{app:B}
Integrating out the vector resonances at tree level, we obtain the effective Lagrangian containing the gauge fields $\bar{W}^\alpha,\bar{B}$ and the Higgs,
\begin{equation}
\mathcal{L}^{\rm eff}_g = \frac{1}{2} \Big(g^{\mu\nu} - \frac{p^\mu p^\nu}{p^2}\Big) \left( 2 \Pi_{+ - } \bar{W}^+_\mu \bar{W}^+_\nu + \Pi_{33} \bar{W}^3_\mu \bar{W}^3_\nu + \Pi_{BB} \bar{B}_\mu \bar{B}_\nu + 2 \Pi_{3B} \bar{W}^3_\mu \bar{B}_\nu\right)
\end{equation}
where
\begin{equation}
\Pi_{+-} = \Pi_{33} = \Pi_0 + \frac{\tilde{h}^2}{4f^2} \Pi_1^g\,,\quad  \Pi_{BB} = \Pi_B + \frac{\bar{g}^{\prime\,2}}{\bar{g}^2} \frac{\tilde{h}^2}{4f^2} \Pi_1^g\,,\quad  \Pi_{3B} = - \frac{\bar{g}'}{\bar{g}} \frac{\tilde{h}^2}{4f^2} \Pi_1^g\,.
\end{equation}
The dynamics of the strong sector resonances are encoded in the momentum-dependent form factors, which read in Euclidean space
\begin{equation}
\Pi_{0 (B)} = p^2 \Bigg(1 + \frac{\bar{g}^{(\prime)\, 2} f_\rho^2}{p^2 + m_\rho^2} \Bigg), \qquad \Pi_1^g = \bar{g}^2 \left[ f^2 + 2 p^2 \left(\frac{f_a^2}{p^2 + m_a^2} - \frac{f_\rho^2}{p^2 + m_\rho^2}\right)\right]. 
\end{equation}
The effective potential for the Higgs has the expression
\begin{equation} \label{eq:Vg}
V_g(\tilde{h}) = \frac{3}{2} \int \frac{d^4 p}{(2\pi)^4} \log \left[ \Pi_{+-}^2 (\Pi_{33} \Pi_{BB} - \Pi_{3B}^2)\right].
\end{equation}

Integrating out the fermionic resonances at tree level we obtain an effective Lagrangian containing the top quark, the $b_L$ and the GBs as degrees of freedom, 
\begin{equation}
\mathcal{L}_t^{\rm eff} = \Pi_{L_0} \bar{b}_L \slashed{p} b_L + \Pi_{L} \bar{t}_L \slashed{p} t_L + \Pi_R \bar{t}_R \slashed{p} t_R -  \left(\Pi_{LR}\, \bar{t}_L t_R + \mathrm{h.c.}\right),
\label{eq:effFermLagr}
\end{equation}
where
\begin{equation} \label{eq:fermiFFs0}
\Pi_L = \Pi_{L_0} + \frac{\tilde{h}^2}{2 f^2}\,\Pi_{L_1}\,, \quad \Pi_R = \Pi_{R_0} + \Bigg(\frac{\tilde{h}^2}{f^2} + \frac{2 \chi^\ast \chi}{f^2}\Bigg) \Pi_{R_1}\,, \quad
\Pi_{LR} = \frac{\tilde{h}}{\sqrt{2} f }\sqrt{1-\frac{\tilde{h}^2}{f^2}-\frac{2 \chi^\ast \chi}{f^2}}\;\Pi_1^t \,.
\end{equation}
The momentum-dependent form factors read, in Euclidean space,
\begin{align}
\Pi_{L_0} = &\; 1+\sum_{i=1}^{N_Q} \frac{\lvert \epsilon_{qQ}^i \rvert^2}{p^2 + m_{Q_i}^2}\,, \qquad \Pi_{L_1}= \sum_{j=1}^{N_S} \frac{\lvert \epsilon_{qS}^j \rvert^2}{p^2 + m_{S_j}^2} - \sum_{i=1}^{N_Q} \frac{\lvert \epsilon_{qQ}^i \rvert^2}{p^2+m_{Q_i}^2} \,,\nonumber \\  \Pi_{R_0} = &\; 1 + \sum_{j=1}^{N_S} \frac{\lvert \epsilon_{tS}^j \rvert^2}{p^2 + m_{S_j}^2}\,, \qquad \Pi_{R_1} = \sum_{i=1}^{N_Q} \frac{\lvert \epsilon_{tQ}^i \rvert^2}{p^2 + m_{Q_i}^2} - \sum_{j=1}^{N_S} \frac{\lvert \epsilon_{tS}^j \rvert^2}{p^2 + m_{S_j}^2} \,, \label{eq:fermiFFs} \\
&\quad\;\; \Pi_1^t = \sum_{j=1}^{N_S} \frac{\epsilon_{tS}^{* j}\epsilon_{qS}^j m_{S_j}}{p^2 + m_{S_j}^2}  - \sum_{i=1}^{N_Q} \frac{\epsilon_{tQ}^{* i} \epsilon_{qQ}^i m_{Q_i}}{p^2 + m_{Q_i}^2} \,. \nonumber
\end{align}
The effective potential for the GBs reads
\begin{equation} \label{eq:Vf}
V_f (\tilde{h},\chi ) = - 2 N_c \int\frac{d^4 p}{(2\pi )^4} \log\left( p^2 \Pi_L \Pi_R  + \lvert \Pi_{LR} \rvert^2 \right).
\end{equation}
Expanding Eqs.~\eqref{eq:Vg} and \eqref{eq:Vf} to quartic order in the fields and matching with Eq.~\eqref{eq:GBPotential}, we obtain the expressions of the parameters $\mu_h^2, \lambda_h, \mu_{\rm DM}^2, \lambda_{\rm DM}$ and $\lambda\,$ as integrals over the form factors. For the dominant fermion contribution we find
\begin{align}
\mu_{h,f}^2 \,=&\, - \frac{N_c}{8\pi^2 f^2} \int_{0}^\infty dp^2 p^2 \left(\frac{\Pi_{L_1}}{\Pi_{L_0}} + \frac{2\Pi_{R_1}}{\Pi_{R_0}} + \frac{(\Pi_1^t)^2}{p^2\Pi_{L_0} \Pi_{R_0}}\right), \nonumber \\
\lambda_{h,f} \,=&\, \frac{N_c}{4\pi^2 f^4} \int_{\mu_{\rm IR}^2}^\infty dp^2 p^2 \left[\frac{1}{4}\left(\frac{\Pi_{L_1}}{\Pi_{L_0}} + \frac{2\Pi_{R_1}}{\Pi_{R_0}} + \frac{(\Pi_1^t)^2}{p^2\Pi_{L_0} \Pi_{R_0}}\right)^2 + \frac{(\Pi_1^t)^2 - p^2 \Pi_{L_1} \Pi_{R_1}}{p^2\Pi_{L_0} \Pi_{R_0}}  \right], \nonumber \\
\mu^2_{\rm DM} \,=&\, - \frac{N_c}{4\pi^2 f^2} \int_{0}^\infty dp^2 p^2 \frac{\Pi_{R_1}}{\Pi_{R_0}}\,,\qquad\quad \lambda_{\rm DM} = \frac{N_c}{4\pi^2 f^4} \int_{0}^\infty dp^2 p^2 \frac{\Pi_{R_1}^2}{\Pi_{R_0}^2}\,, \nonumber \\
\lambda \,=&\, \frac{N_c}{8\pi^2 f^4} \int_{0}^\infty dp^2 p^2 \left[2\,\frac{\Pi_{R_1}^2}{\Pi_{R_0}^2} + \frac{(\Pi_1^t)^2}{p^2 \Pi_{L_0} \Pi_{R_0}} \left(1 + \frac{\Pi_{R_1}}{\Pi_{R_0}}\right)\right], \label{eq:FFs}
\end{align}
where we assumed real mixing parameters $\epsilon$. Notice that the integral for the Higgs quartic $\lambda_{h,f}$ is IR divergent; the same happens for the (small) gauge contribution $\lambda_{h,g}$. The IR divergence signals that the potential is non-analytic at $\tilde{h} = 0$, due to the contribution of the light degrees of freedom (the top quark and SM gauge bosons). To remove this issue, the expansion of the potential in Eq.~\eqref{eq:GBPotential} is extended to include an additional term \mbox{$\Delta V = (\delta_h/2) \tilde{h}^4 \log (\tilde{h}^2/f^2)$}, which captures the non-analytic contribution to the Higgs quartic. Then all the coefficients of $V + \Delta V$ are IR-finite, including $\delta_h$. The Higgs VEV \mbox{$\langle \tilde{h} \rangle = v$} is obtained by solving the equation $\langle \tilde{h} \rangle^2 = - \mu_h^2/ [\lambda_h + \delta_h(1 + 2 \log (\langle \tilde{h} \rangle^2/f^2))]$, and the Higgs mass is $m_h^2 = (1 - \xi) 2 v^2 (\lambda_h + 3\delta_h + 2 \delta_h \log \xi)$.

We now summarize our procedure for the parameter scan. From Eq.~\eqref{eq:angles2layers}, requiring that $0 \leq s_{\theta, \phi}^2 \leq 1$ leads to the constraints
\begin{equation} \label{eq:anglebound}
\frac{m_{S_1}^2 - m_{Q_1}^2}{m_{Q_2}^2 - m_{Q_1}^2} \leq s_{\alpha,\beta}^2 \leq \frac{m_{S_2}^2 - m_{Q_1}^2}{m_{Q_2}^2 - m_{Q_1}^2}\,.
\end{equation}
These can be satisfied only for $m_{S_2} > m_{Q_1}$, which we therefore assume. Taking into account that $\Pi_1^t$ is the only form factor that is sensitive to the signs of the mixing parameters $\epsilon$, and that furthermore the scalar potential is unaffected by $\Pi_1^t \to - \Pi_1^t$, the angles are restricted to the following ranges 
\begin{equation} \label{eq:angleranges}
\theta, \alpha \in [-\pi/2, \pi/2],\qquad \phi \in [0, \pi/2],\qquad \beta \in [0, \pi].
\end{equation}
We summarize here the procedure adopted in the parameter scan of the two-layer model with WSRs (the procedure for the scan of the one-layer model is analogous).
\begin{enumerate}
\item The following parameters are randomly selected: $\epsilon_t \in \left[f/10, 8f\right],\quad m_{S_1, Q_1}\in \left[0,6f\right]\,$, \mbox{$m_{S_2 , Q_2}\in\left[m_{Q_1}, 6f\right]$}, $f_\rho \in \left [f / \sqrt{2}, 2 f\right]\,$;
\item The angles $\alpha$ and $\beta$ are randomly picked, compatibly with the restrictions in Eqs.~\eqref{eq:anglebound} and \eqref{eq:angleranges}. Then $\phi$ is completely fixed, while the sign of $\sin\theta$ is picked randomly.
\item $\epsilon_q$ is fixed by solving the following equation
\begin{equation}
m_t^2 = \frac{\left| \Pi_{LR} (m_t^2)\right|^2 }{\Pi_L(m_t^2) \Pi_R(m_t^2)} \Bigg|_{\,\tilde{h}\, =\, v,\, \chi\, =\, 0}\,,
\end{equation}
where the numerical value of the top mass is set to $m_t = m_t^{\overline{MS}}(2\; \mathrm{TeV}) = 150$ GeV.
\item $m_\rho$ is fixed by requiring the Higgs VEV to match the observed value, $\langle \tilde{h} \rangle = v \simeq 246\;\mathrm{GeV}$.
\end{enumerate}

In the two-layer model, the compositeness fraction $s_L\,(s_R)$ of the left (right) handed top is computed by diagonalizing analytically the fermion mass matrix for $v\to 0$, and taking the projection onto the composite fermions of the normalized eigenvector that corresponds to the physical $t_L\,(t_R)$. For example, the compositeness fraction of $t_R$ is defined as
\begin{equation} \label{eq:tRcmpfraction}
s_R \equiv \sqrt{ \frac{a_2^2 + a_3^2}{a_1^2 + a_2^2 + a_3^2}  }\, \qquad \mathrm{with}\qquad t_R^{p} = \frac{1}{\sqrt{a_1^2 + a_2^2 + a_3^2}} \,(a_1 t_R + a_2 S_1 + a_3 S_2),
\end{equation}  
where $t_R^{p}$ denotes the mass-eigenstate right-handed top (for $v\to 0$). The compositeness fractions satisfy $0 \leq s_{L,R} \leq 1$. In the one-layer model, they are identified with the sine of the elementary-composite mixing angles.

\section{Details on DM phenomenology} \label{app:C}
The explicit values of the couplings in Eqs.~(\ref{eq:LEGBint},\,\ref{eq:Veff}) are
\begin{equation}
\begin{split} \label{eq:EFTcouplingsANN}
a_{hhh}& = b_{h\chi\chi} = \frac{\xi}{\sqrt{1-\xi}} \,,\qquad a_{hh\chi\chi}=\frac{\xi^2}{1-\xi}\,,\qquad b_{hh\chi\chi}=\xi \,\frac{1+\xi}{1-\xi}\,,\\
&a_{hVV} = d_{h\chi\chi} = d_{hhh} = \sqrt{1-\xi}\,, \qquad d_{hh\chi\chi}=1-\xi\,.
\end{split}
\end{equation}
The couplings between the scalars and the top quark in Eq.~\eqref{eq:topEffLagr} can be easily computed by matching with Eq.~\eqref{eq:effFermLagr}, where the top partners have been integrated out in the original field basis. However, the results of the parameter scan show that the ``composite'' mass of the lightest singlet, $m_{S_1}$, can in some cases be as low as few hundred GeV (while the physical mass of the lightest singlet is still above the experimental lower bound of $1$ TeV, because it receives a large contribution from the elementary-composite mixing parameters $\sim \epsilon_t$), thus invalidating the simple effective theory approach in this basis. Therefore we proceed as follows: Starting from the UV Lagrangian in Eq.~\eqref{eq:fermion}, after exact, numerical diagonalization of the fermion mass matrices we consider the following terms
\begin{align} \label{eq:UVlagr}
\mathcal{L}_f\;  \ni \; &\;i\bar{t}\slashed{\partial}t -m_t \bar{t}t\Big( \tilde{c}_{tth} \frac{h}{v} + 2\, \tilde{c}_{tt\chi\chi}\frac{\chi^\ast \chi}{v^2}\Big) \\ 
+\;& \sum_{i=1}^{N_Q} \Big[\overline{\mathcal{Y}}_i\left(i\slashed{\partial}-m_{Q_i}\right)\mathcal{Y}_i + \overline{\mathcal{Z}}_i\left(i\slashed{\partial}-m_{Q_i}\right)\mathcal{Z}_i
+  \bar{t} ( b^{i}_L P_L + b^{i}_R P_R ) (\mathcal{Y}_i \chi^\ast + \mathcal{Z}_i \chi ) + \mathrm{h.c.} \Big], \nonumber
\end{align}
where we introduced the coefficients $\tilde{c}_{tth}, \tilde{c}_{tt\chi\chi}, b_L^i$ and $b_R^i$, which are real if $CP$ invariance is imposed. After integrating out the $\mathcal{Y}_i$ and $\mathcal{Z}_i$ and matching to Eq.~\eqref{eq:topEffLagr}, we find that $c_{tth} = \tilde{c}_{tth}$, whereas
\begin{equation}
c_{tt\chi\chi} = \tilde{c}_{tt\chi\chi} - \frac{v^2}{m_t} \sum_{i=1}^{N_Q} \Bigg[\frac{b^{i}_L b^{i}_R}{m_{Q_{i}}}+\frac{m_t}{2 m_{Q_i}^2}(b^{i\,2}_L + b^{i\,2}_R)\Bigg].
\end{equation}
We have verified that for parameter choices where the EFT approximation is justified, the values of $c_{tth}$ and $c_{tt\chi\chi}$ obtained from Eq.~\eqref{eq:effFermLagr} agree with those computed with this semi-numerical method.

The cosmological evolution of the $\chi$ number density\footnote{Notice that the DM number density is obtained summing over particles and anti-particles, $n_{\rm DM}=2\,n_\chi$.} is described by the Boltzmann equation
\begin{equation} \label{eq:Boltzmann}
\frac{d n_\chi}{dt} + 3 H n_\chi =  - \langle \sigma v_{\text{rel}} \rangle \left[n_\chi^2-\left(n_\chi^{\rm eq}\right)^2\right],
\end{equation}
where $n_\chi^{\rm eq}$ is the equilibrium number density, $H$ is the time-dependent Hubble parameter and $\langle \sigma v_{\text{rel}} \rangle$ is the thermally averaged annihilation cross-section times the relative velocity of two DM particles, whose expression is \cite{Gondolo:1990dk}
\begin{equation}
\langle \sigma v_{\text{rel}} \rangle (T) = \frac{1}{16 m_\chi^4 T K_2^2(m_\chi /T)}\int_{4m_\chi^2}^\infty ds\, s\, \sqrt{s-4m_\chi^2}\, K_1 (\sqrt{s}/T)\, \sigma v_{\rm rel}(s)\,,
\end{equation}
where $T$ denotes the temperature and $K_1,K_2$ are modified Bessel functions of the second kind. Dark matter annihilates dominantly into $WW, ZZ, hh$ and $t\bar{t}$. The corresponding cross sections were calculated analytically in terms of the parameters of the effective Lagrangian in Eq.~\eqref{eq:Leff_ann}, and found to agree with those of Ref.~\cite{Frigerio:2012uc} in the limit $c_{tth} = c_{tth}^{\rm nl\sigma m}$, \mbox{$c_{tt\chi\chi} = c_{tt\chi \chi}^{\rm nl\sigma m}$}. Equation \eqref{eq:DMrelic} provides a naive solution of the Boltzmann equation, which is nevertheless useful for a qualitative understanding.

The leading $1$-loop corrections to the derivative $\chi \chi^\ast hh$ coupling in Eq.~\eqref{eq:derivtree} are obtained computing the set of Feynman diagrams depicted in Fig.~\ref{fig:FiniteMomCorrections}, and selecting the logarithmically divergent pieces. For simplicity, we report the result in the limit where the GB masses are neglected. Even though this is a rough approximation for DM annihilation, where the kinematic variables take the values (assuming $m_\chi^2 \gg m_h^2$) $s \sim 4m_\chi^2$ and $t \sim u \sim - m_\chi^2$, it is nevertheless sufficient for the purpose of estimating the theoretical uncertainty on the cross section. In particular, it implies that $s + t + u \simeq 0$. The first class of diagrams in Fig.~\ref{fig:FiniteMomCorrections}, which contain two insertions of the elementary-composite mixings, yield the result in momentum space
\begin{equation} \label{eq:class1}
\frac{i N_c}{8\pi^2 f^4} \Big(\epsilon_t^2 - \frac{\epsilon_q^2}{8}\Big) s \log \Lambda^2\,. 
\end{equation}
Notice that this class of diagrams also yield the $O(p^0)$ coupling $\lambda$. The second class of diagrams contain two derivative couplings arising from the $e_\mu$ symbol, and give
\begin{equation} \label{eq:class2}
\frac{i N_c}{8\pi^2 f^4} \Big( - \frac{\epsilon_q^2}{8}\Big) 3\,s \log \Lambda^2\,,
\end{equation}
which can be seen as arising from two $\epsilon$ insertions on the internal fermion lines. The contribution of the third class of diagrams turns out to be proportional to the external masses, and thus negligible within our approximations. Lastly, the triangle diagrams composing the fourth class yield
\begin{equation} \label{eq:class4}
\frac{i N_c}{8\pi^2 f^4} \big( \epsilon_t^2 \big) 3\,s \log \Lambda^2\,. 
\end{equation}
Summing Eqs.~\eqref{eq:class1}, \eqref{eq:class2} and \eqref{eq:class4} and making the argument of the logarithm dimensionless by inserting $m_\ast^2$, we arrive at the final result in Eq.~\eqref{eq:finitemomResult}.

The SI DM-nucleon cross section is given by
\begin{equation}
\sigma_{\rm SI}^{\chi N} = \frac{1}{\pi}\bigg(\frac{m_N}{m_\chi + m_N}\bigg)^2 \left[\frac{Z F_p + (A - Z) F_n}{A}\right]^2,
\end{equation}
where $m_N = (m_p + m_n)/2$ is the average nucleon mass, and for Xenon $A = 130$, $Z = 54$. The effective couplings of the DM to nucleons can be written as
\begin{equation} \label{eq:fN}
\frac{F_x}{m_x} = \sum_{q = u,d,s} f_{T_q}^x a_q + \frac{2}{27}\, f_{T_g}^x \bigg( \sum_{q = c,b} a_q + k_g^t\bigg),\qquad (x = p,\,n)
\end{equation} 
where the first term represents the tree-level coupling to the light quarks $u,d,s$, while the second term parameterizes the coupling to gluons via loops of heavy fermions. For convenience, in the second term we have further singled out the contribution mediated by the top and top partners, $k_g^t$, from the one coming from the charm and bottom. The former can be easily computed using the low-energy theorem for the GBs,
\begin{equation} \label{eq:kgt}
k_g^t = \frac{\lambda v}{m_h^2}\, d_{h\chi\chi} D_h - \frac{1}{2} D_{\chi\chi^\ast}
\end{equation}
with the definitions
\begin{equation}
\begin{split} \label{eq:Ds}
D_h &\equiv \sqrt{1 - \xi}\,\left(\frac{\partial}{\partial \tilde{h}} \log | \mathrm{det}\, \mathcal{M}_t (\tilde{h}, \chi) | \right)_{\tilde{h}\, =\, v,\, \chi\,  =\, 0} \;=\; \frac{1}{v} \frac{1-2\xi}{\sqrt{1-\xi}}\,,\\ 
D_{\chi \chi*} &\equiv \left(\frac{\partial^2}{\partial \chi \partial \chi^\ast} \log | \mathrm{det}\, \mathcal{M}_t (\tilde{h}, \chi) | \right)_{ \tilde{h}\,  =\, v,\,  \chi \, =\, 0} \;=\; -\, \frac{1}{f^2 (1 - \xi)}\,,
\end{split}
\end{equation}
where $\mathcal{M}_t$ is the field-dependent mass matrix for the top sector. Even though $k_g^t$ receives contributions from the top partners, its final expression depends only on $f$ and is insensitive to the resonance parameters. This cancellation can be traced to the fact that with our choice of fermion embeddings, $q_L, t_R \sim \mathbf{7}$ of $SO(7)$, there is only one $SO(6)$ invariant that generates the top mass \cite{Azatov:2011qy,Montull:2013mla}.\footnote{Notice that the expression of $k_g^t$ is identical to that obtained in $SO(6)/SO(5)$ when $q_L, t_R \sim \mathbf{6}$ \cite{Frigerio:2012uc}.} We remark that our computation based on Eq.~\eqref{eq:kgt} is only approximate for the box diagrams that contain $\mathcal{Y}, \mathcal{Z}$ propagators, and could be improved through an exact computation of the $\chi g \to \chi g$ scattering amplitude, see Ref.~\cite{Drees:1993bu} for an extensive discussion in the similar case of neutralino-nucleon scattering. However, we have checked that for realistic parameter points the contribution of the box diagrams to $k_g^t$ is $\lesssim 10\%$, hence we estimate that the corrections to our approximation would only affect $\sigma_{\rm SI}^{\chi N}$ at the percent level. 

The contribution of the light SM quarks is encoded by the coefficients $a_q$ ($q = u,d,c,s,b$) in Eq.~\eqref{eq:fN}. It is somewhat model-dependent, being determined by the choice of the corresponding embeddings, which we have not specified so far since they do not affect any other aspect of the phenomenology. For concreteness, we assume all left-handed light quarks to be embedded in the $\mathbf{7}$, whereas for the right-handed light quarks we take $b_R\sim \mathbf{7}$, leading to a contribution identical to the one of the top sector, and $q_R \sim \mathbf{1}$ ($q = u, d, c, s$), yielding a vanishing coefficient for the $\chi^\ast \chi \bar{q} q$ contact term. In summary, we have
\begin{equation}
k_g^t = a_b = \frac{\lambda}{m_h^2} ( 1 - 2\xi ) + \frac{1}{2 f^2 (1 - \xi)}\,, \qquad a_{u,d,c,s} = \frac{\lambda}{m_h^2} ( 1 - \xi ).
\end{equation} 

For the nuclear matrix elements that appear in Eq.~\eqref{eq:fN} we take $f_{T_u}^p =  0.021$, \mbox{$f_{T_d}^p = 0.041$}, $f_{T_u}^n = 0.019$, $f_{T_d}^n = 0.045$, obtained from agreeing determinations of the pion-nucleon sigma term $\sigma_{\pi N}$ from chiral perturbation theory~\cite{Alarcon:2011zs} and dispersive methods~\cite{Hoferichter:2015dsa}, and $f_{T_s}^{p,n} = 0.043$, based on lattice QCD results~\cite{Junnarkar:2013ac}. The gluon matrix element is then $f_{T_g}^{p,n} = 1 - \sum_{q = u,d,s} f_{T_q}^{p,n} \simeq 0.89$. For realistic parameters the Higgs exchange dominates and the cross section can be approximated by the simple expression in Eq.~\eqref{eq:DDxsection}, with $f_N \equiv 2/9 + (7/9) \sum_{q = u,d,s} f_{T_q}^{p,n} \simeq 0.30$.

%
\end{document}